\documentclass[10pt]{article}

\usepackage[preprint]{tmlr}     %

\usepackage[utf8]{inputenc}
\usepackage{amsmath,amssymb,amsthm}
\usepackage{booktabs}
\usepackage{array}   %
\usepackage{longtable}  %
\usepackage{graphicx}
\usepackage{microtype}
\usepackage[hidelinks]{hyperref}
\usepackage{xcolor}
\usepackage{caption}
\setcitestyle{aysep={}}

\makeatletter
\if@accepted
  \newcommand{\deanon}[2]{#1}
\else
  \newcommand{\deanon}[2]{#2}
\fi
\makeatother

\newcommand{\atCells}{12}

\newcommand{\atDefaultGot}{0.6484}

\newcommand{\atDefaultNominal}{0.90}

\newcommand{\atSeries}{1479}
\newcommand{\atStrict}{11}

\newcommand{\bsCells}{24}
\newcommand{\bsCellsGapBelowSd}{21}
\newcommand{\bsGapFifty}{0.0196}

\newcommand{\bsPBelowFifty}{0.4312}
\newcommand{\bsPBelowShortFifty}{0.6161}

\newcommand{\bsRatioThousand}{0.1055}
\newcommand{\bsSdFifty}{0.0412}

\newcommand{\ceCertSize}{4}
\newcommand{\ceClasses}{11}

\newcommand{\ceLevels}{3}
\newcommand{\ceRules}{16}

\newcommand{\cenAbsent}{13}
\newcommand{\cenCorrected}{1}
\newcommand{\cenDriven}{12}

\newcommand{\cenExcluded}{2}
\newcommand{\cenExecuted}{23}
\newcommand{\cenLibraries}{10}

\newcommand{\cenPTwo}{39}
\newcommand{\cenPThree}{38}
\newcommand{\cenPvalue}{3}
\newcommand{\cenRead}{13}
\newcommand{\cenShared}{11}

\newcommand{\clVerNumpy}{2.4.6}

\newcommand{\clipCells}{24}
\newcommand{\clipDelivered}{0.9091}
\newcommand{\clipLevel}{0.95}
\newcommand{\clipLevels}{2}
\newcommand{\clipMaxZ}{1.19}
\newcommand{\clipMeasured}{0.9078}

\newcommand{\clipN}{10}
\newcommand{\clipPublic}{6}
\newcommand{\clipReaches}{3}
\newcommand{\clipReps}{50000}
\newcommand{\clipReq}{11}
\newcommand{\clipSE}{0.0013}
\newcommand{\clipShortfall}{0.0409}
\newcommand{\clipSigs}{8}
\newcommand{\clipTried}{3}

\newcommand{\clipWindowTop}{18}
\newcommand{\clipWorstDelivered}{0.6667}

\newcommand{\clipWorstN}{2}
\newcommand{\clipWorstShortfall}{0.2833}
\newcommand{\confBelowCount}{1}
\newcommand{\confBelowFinite}{8}
\newcommand{\confBelowInf}{3}
\newcommand{\confBelowRaise}{2}
\newcommand{\confBelowSilent}{9}

\newcommand{\confPaths}{14}
\newcommand{\confRouteInf}{3}
\newcommand{\confRouteRaise}{6}
\newcommand{\confRouteSilent}{5}
\newcommand{\confSharedPaths}{1}
\newcommand{\confSites}{13}
\newcommand{\confWarns}{3}

\newcommand{\daCoincCells}{3}
\newcommand{\daDeficCells}{3}
\newcommand{\daDeficMax}{0.0560}
\newcommand{\daDeficMin}{0.0280}
\newcommand{\daMThreeDeficMax}{0.0640}
\newcommand{\daMThreeDeficMin}{0.0120}

\newcommand{\daNinetyOneFiveD}{0.0560}

\newcommand{\daNinetyOneFiveDisc}{14}

\newcommand{\daWidthRatioMean}{1.3684}

\newcommand{\dmOne}{40}
\newcommand{\dmZero}{12}
\newcommand{\dsAgeMax}{129}
\newcommand{\dsAsOf}{2026-11-16}
\newcommand{\dsClosed}{4}
\newcommand{\dsEngaged}{11}

\newcommand{\dsFilerOnly}{3}
\newcommand{\dsFilings}{16}

\newcommand{\dsLinkedFixPRs}{3}

\newcommand{\dsMerged}{6}
\newcommand{\dsOpen}{9}

\newcommand{\dsSilent}{5}
\newcommand{\dsTargets}{19}

\newcommand{\exactZero}{0.0000}

\newcommand{\fmDeficitMean}{0.0293}

\newcommand{\fsDate}{2026-09-01}

\newcommand{\fsOutside}{36}

\newcommand{\fsQcCells}{55}

\newcommand{\glCells}{28}

\newcommand{\glOffCells}{28}
\newcommand{\glOffWorst}{0.00059}

\newcommand{\glWorstLaw}{0.000542}

\newcommand{\hzAlignReps}{300}
\newcommand{\hzCells}{14}

\newcommand{\hzFloorFiftyTwoNinetyFive}{1039}

\newcommand{\hzFloorTwelveNinetyFive}{239}
\newcommand{\hzFloorTwentyFourNinetyFive}{479}
\newcommand{\hzInfeasCells}{7}

\newcommand{\hzOneA}{0.9667}

\newcommand{\hzOneD}{0.8867}

\newcommand{\hzRankCost}{0.0900}
\newcommand{\hzRatioFiftyTwo}{54.7}
\newcommand{\hzRatioTwelve}{12.6}

\newcommand{\hzVarOne}{1.899}
\newcommand{\hzVarOneExp}{1.919}
\newcommand{\hzVarOneSe}{0.028}
\newcommand{\hzVarTwo}{2.782}
\newcommand{\hzVarTwoExp}{2.816}
\newcommand{\hzVarTwoSe}{0.048}
\newcommand{\hzVarZero}{0.978}
\newcommand{\hzVarZeroExp}{0.980}
\newcommand{\hzVarZeroSe}{0.012}

\newcommand{\iwAudTotal}{4,061,258}
\newcommand{\iwAudited}{10}
\newcommand{\iwCfAudTotal}{245,601}
\newcommand{\iwCfShare}{81.6}
\newcommand{\iwCfUnaudTotal}{55,535}
\newcommand{\iwDate}{2026-08-06}
\newcommand{\iwShare}{20.1}
\newcommand{\iwUnaudTotal}{16,116,290}
\newcommand{\iwUnaudited}{12}
\newcommand{\lacCountAtBad}{8.1}

\newcommand{\muCells}{56}

\newcommand{\muDropped}{1}
\newcommand{\muDroppedDelta}{0.0120}
\newcommand{\muFamily}{8}
\newcommand{\muMaxFactor}{2}
\newcommand{\muSurvive}{7}
\newcommand{\muWorstLevel}{0.999}
\newcommand{\muWorstN}{1000}

\newcommand{\oosCtCells}{36}
\newcommand{\oosCtClaim}{24}
\newcommand{\oosCtClipped}{20}
\newcommand{\oosCtDecisive}{15}
\newcommand{\oosCtFloor}{128}
\newcommand{\oosCtHeadLevel}{0.975}
\newcommand{\oosCtHeadN}{135}
\newcommand{\oosCtHeadRequired}{133}
\newcommand{\oosCtHeadVirtual}{131.650}
\newcommand{\oosCtSkip}{90}
\newcommand{\oosCtUnder}{24}
\newcommand{\oosCtUnderClipped}{9}
\newcommand{\oosCwClassicNmin}{10}
\newcommand{\oosCwClassicRank}{37.000}
\newcommand{\oosCwWeightedNmin}{20}
\newcommand{\oosCwWeightedRank}{38.000}
\newcommand{\oosFortunaBad}{8.000}
\newcommand{\oosFortunaBranch}{b}
\newcommand{\oosFortunaBranchName}{clamps}

\newcommand{\oosFortunaRank}{37.075}
\newcommand{\oosNbad}{8}
\newcommand{\oosNok}{40}
\newcommand{\oosNone}{6}
\newcommand{\oosPackages}{3}
\newcommand{\oosRefuted}{2}
\newcommand{\oosRequired}{37}
\newcommand{\oosSearched}{12}
\newcommand{\oosSite}{4}
\newcommand{\oosSkforecastBad}{7.300}
\newcommand{\oosSkforecastBranch}{d}
\newcommand{\oosSkforecastRank}{36.100}
\newcommand{\pairedCells}{148}
\newcommand{\pairedDaCells}{24}
\newcommand{\pairedLossCells}{6}
\newcommand{\pairedLossConfigs}{4}
\newcommand{\pairedReversals}{80}
\newcommand{\pairedSfCells}{40}
\newcommand{\pairedSkCells}{16}
\newcommand{\pairedTabCells}{68}
\newcommand{\pairedUnits}{20961}

\newcommand{\pfAppendixPage}{16}

\newcommand{\pfBodyPage}{7}

\newcommand{\rcCeilDefs}{13}

\newcommand{\rcCeilExactNinety}{11}

\newcommand{\rcCeilHigherOverNinety}{382}
\newcommand{\rcCeilInexactNinety}{373}

\newcommand{\rcCeilShortNinety}{3}
\newcommand{\rcCeilSizesNinety}{392}

\newcommand{\rcExecCells}{1941}

\newcommand{\rcExecMiss}{21}

\newcommand{\rcHigherExactNinety}{351}

\newcommand{\rcHigherSizesNinety}{390}

\newcommand{\rcLevels}{6}

\newcommand{\rfcBelowFloor}{48}

\newcommand{\rfcFixtures}{51}

\newcommand{\rmDefs}{13}
\newcommand{\rmDeliverNinety}{4}

\newcommand{\rmTheoreticalNinety}{9}
\newcommand{\rmTheoreticalNinetyFive}{19}
\newcommand{\robArchive}{1428}

\newcommand{\robCompared}{28}

\newcommand{\robMOneSkBeyond}{0}
\newcommand{\robMOneSkCells}{8}

\newcommand{\robOrigins}{4}

\newcommand{\robSeries}{1093}

\newcommand{\robSfDefaultNinety}{0.6128}

\newcommand{\robSignAgree}{28}
\newcommand{\robSkCells}{8}

\newcommand{\rqOneContradicted}{3}
\newcommand{\rqOneResisted}{4}
\newcommand{\rvCovFifty}{0.8732}
\newcommand{\rvCovTen}{0.4505}
\newcommand{\rvCovThousand}{0.9022}
\newcommand{\rvCovThree}{0.4795}
\newcommand{\rvCovTwoHundred}{0.8935}
\newcommand{\rvReps}{4000}
\newcommand{\rvRequested}{0.90}
\newcommand{\rvVersion}{0.25.0}
\newcommand{\scAddedAfter}{3}

\newcommand{\scCorrect}{11}
\newcommand{\scDrawn}{12}
\newcommand{\scKappa}{1.000}

\newcommand{\scScoreable}{11}

\newcommand{\sfCutKstarTwoZero}{19}

\newcommand{\sfCutPosTwoZero}{18.05}

\newcommand{\sfDefaultNinety}{0.5840}

\newcommand{\sfDefaultNinetyPredA}{0.6169}

\newcommand{\sfDistNinetyOneZeroD}{0.1080}
\newcommand{\sfDistNinetyOneZeroPred}{0.0857}

\newcommand{\sfDistNinetyTwoZeroA}{0.8640}
\newcommand{\sfDistNinetyTwoZeroD}{0.0400}
\newcommand{\sfDistNinetyTwoZeroPred}{0.0434}

\newcommand{\sfDistNinetyTwoZeroSE}{0.0124}

\newcommand{\sfDistNinetyFiveZeroD}{0.0120}
\newcommand{\sfDistNinetyFiveZeroPred}{0.0178}

\newcommand{\sfErrDefaultNinety}{0.6040}
\newcommand{\sfErrDefaultNinetyFive}{0.6080}

\newcommand{\sfGapTwoZero}{10}

\newcommand{\sfLedgerUnits}{250}
\newcommand{\sfMThreeDefaultNinety}{0.6120}

\newcommand{\sfPosMeanTwoZero}{18.088}

\newcommand{\sfSweepFeasible}{392}
\newcommand{\sfSweepMaxShort}{2}
\newcommand{\sfSweepShortOne}{80}
\newcommand{\sfSweepShortTwo}{312}

\newcommand{\sfSweepTo}{400}
\newcommand{\sfZWorst}{1.14}

\newcommand{\skCells}{8}
\newcommand{\skConfNinetyTwentyA}{0.9000}
\newcommand{\skEmpNinetyFortyA}{0.8880}
\newcommand{\skEmpNinetyFortyD}{0.0640}

\newcommand{\skEmpNinetyTwentyA}{0.8920}

\newcommand{\skEmpNinetyTwentyD}{0.0360}
\newcommand{\skEmpNinetyTwentyDisc}{9}
\newcommand{\skEmpNinetyTwentyLoss}{0}
\newcommand{\skEmpNinetyTwentyN}{74}
\newcommand{\skEmpNinetyTwentyRank}{65}
\newcommand{\skEmpNinetyTwentyReq}{68}

\newcommand{\skSeries}{250}

\newcommand{\tabConfigs}{7}
\newcommand{\tabDatasetsClf}{15}
\newcommand{\tabDatasetsReg}{14}

\newcommand{\tabExactConfigs}{3}
\newcommand{\tabExcCells}{8}
\newcommand{\tabExcExact}{5}
\newcommand{\tabExcExactSizes}{20, 200}
\newcommand{\tabExcNonzero}{3}
\newcommand{\tabExcWideSizes}{30, 49, 50}
\newcommand{\tabLitCells}{10}
\newcommand{\tabLitOn}{10}
\newcommand{\tabMaxFeatures}{100}
\newcommand{\tabMaxRows}{5000}

\newcommand{\tabSizes}{20, 30, 49, 50, 200}
\newcommand{\tabSkipClf}{12}
\newcommand{\tabSkipReg}{21}
\newcommand{\tabWorstDelta}{-0.0284}

\newcommand{\tabWorstZ}{4.7}

\newcommand{\wdCells}{30}

\newcommand{\wdCorrMean}{1.1549}
\newcommand{\wdCorrMeanSe}{0.00041}
\newcommand{\wdCorrWorst}{2.5742}

\newcommand{\wdOffCells}{20}
\newcommand{\wdOffCorrMean}{1.1151}

\makeatletter
\def\doi{\begingroup\catcode`\_=12 \catcode`\#=12 \catcode`\&=12 \doi@}
\def\doi@#1{\href{https://doi.org/#1}{\nolinkurl{https://doi.org/#1}}\endgroup}
\makeatother

\newcommand{\orone}{Online Resource~1}
\newcommand{\orsec}[1]{\orone, \S#1}

\newcommand{\code}[1]{\texttt{#1}}
\newtheorem{proposition}{Proposition}
\newcommand{\reqrank}{k^{\star}}
\newcommand{\delrank}{\hat{k}}

\title{A level is not a rank: an empirical audit of quantile-to-index\\
conversion in \cenLibraries{} released conformal-prediction libraries}

\author{\name Shivam Lalakiya \email shivamlalakiya151299@gmail.com \\
        \addr Independent Researcher, St.\ Louis, Missouri, USA \\
        \addr ORCID: 0009-0000-5110-6540}

\begin{document}
\maketitle

\begin{abstract}
\textbf{Context.} A conformal interval at $0.90$ keeps that guarantee only at the right
rank: a fresh score falls at or below the $r$th smallest of $n$ with
probability $r/(n+1)$. The libraries we study instead hand a level to a quantile function,
which can land elsewhere. At $n = 20$, the upper end of \code{statsforecast}'s interval sits
at position $\sfCutPosTwoZero{}$ where rank $\sfCutKstarTwoZero{}$ is required.

\textbf{Objective.} Where does installed code resolve the wrong rank, at what cost in
coverage, and what does it return when no rank reaches the level?

\textbf{Method.} We audit \cenPTwo{} sites where a level becomes an index in
\cenLibraries{} pinned packages, a named set rather than a systematic search. \cenDriven{}
are classified by running them, \cenShared{} through their shared \code{numpy} method,
the rest from source. Coverage is paired against the required rank on the same scores.

\textbf{Results.} All \tabConfigs{} tabular implementations reach the required rank. In
\code{darts}, on one code path, coverage is unchanged (\exactZero{}) where the ranks agree
and \daDeficMin{} to \daDeficMax{} lower where they differ; a second archive repeats that
split but not its range. Below the feasibility floor, $n < 1/\alpha - 1$, of
\confPaths{} public call paths \confRouteSilent{} silently return a finite number; the
rest raise or warn.

\textbf{Conclusions.} Quantile interfaces take only a level;
\code{statsforecast}'s default calibrates below its floor. \dsMerged{} of
\dsTargets{} upstream filings have merged, one fixing a coverage shortfall. A conformance
suite classifies helpers by the rank they reach, and a checklist guides review.
\end{abstract}

\vspace{0.5em}

\vspace{0.5em}
\section{Introduction}
\label{sec:intro}

A practitioner who wants an interval with a stated coverage installs a library, passes a
level, and gets back a threshold. Coverage is the fraction of fresh cases the interval
contains. The threshold comes from a \emph{calibration set}: labelled cases the model did
not train on, one score each, measuring how far that case fell from its prediction.
Sorting those $n$ scores gives the \emph{order statistics}, the 1st through the $n$th
smallest, and the guarantee the practitioner was promised is a statement about exactly one
of them. This construction, a model fitted once and a threshold read off cases it never
saw, is split conformal prediction; the split is between the cases that trained the model
and the cases that calibrate it. The reason the guarantee names one order statistic is
short. Sorted, $n$ scores leave $n+1$ gaps for a fresh score to fall into, and each
gap is equally likely. That is what \emph{exchangeable} buys: the fresh score's joint law
with the calibration scores is unchanged by reordering them, as when all are drawn
independently from one distribution. So the fresh score sits at or below the $r$th
smallest with probability $r/(n+1)$. Take as threshold the $\reqrank$-th smallest score,
$\reqrank = \lceil (n+1)(1-\alpha) \rceil$ being the smallest rank whose
$\reqrank/(n+1)$ reaches the requested $1-\alpha$. The interval it bounds then covers
with probability at least $1-\alpha$, an inequality rather than an equality because
$(n+1)(1-\alpha)$ is rarely an integer. The index is not folklore: it is written out in
the appendix of TorchCP's peer-reviewed description \citep{torchcp}, one of the libraries
we audit.

The difficulty is that no widely used quantile function accepts that rank.
\code{numpy.quantile}, \code{torch.quantile}, and their equivalents accept a
level in $[0,1]$ and return a value whose position among the order statistics
depends on an interpolation convention. \code{numpy}'s default, \code{linear},
maps a level $q$ over $n$ points to the \emph{virtual} index $1 + q(n-1)$ in this
paper's 1-based indexing (\code{numpy}'s documentation writes it 0-based as
$q(n-1)$), which is generally not an integer; the returned number then lies strictly
between two order statistics and no exchangeability argument applies to it directly. A
library author who wants rank $\reqrank$ must convert it to a level, and that conversion
is where coverage is lost.

\code{darts} rounds the requested coverage $c$ straight to an index, so the rank it
lands on is $\lceil c(n-1) \rceil + 1$ where the request needs $\lceil c(n+1) \rceil$.
At some calibration sizes the two coincide and at the rest they differ by one, depending
on $n$ alone. We sampled both bands at a requested $0.90$ on the same library, the same
archive and the same code path, three cells each. The paired difference against an arm
built at the required rank is exactly \exactZero{} on the \daCoincCells{} coincidence
cells and between \daDeficMin{} and \daDeficMax{} on the \daDeficCells{} deficit cells.
On a second archive the same cells split the same way, zero and positive, though not
inside that range (\S\ref{sec:severity}). The six cells differ in $n$, and so in fit
window and scores; within a cell the two arms share the scores and differ in the rank
alone. The zeros are arithmetic coincidence, not correctness, and a study that samples
only them reports a library as correct.

\begin{figure}[t]
\centering
\includegraphics[width=0.86\textwidth]{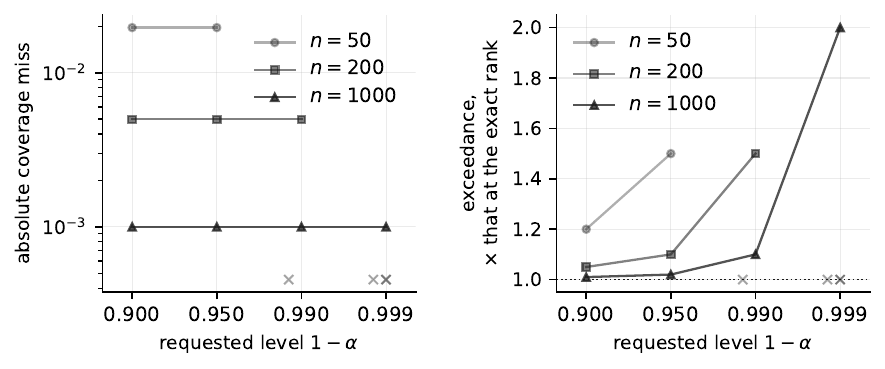}
\caption{The same one-rank deficit, a helper landing one rank below $\reqrank$, drawn
twice. Left: the absolute coverage miss, $1/(n+1)$, flat in the requested level by
construction. Right: the delivered exceedance as a multiple of the exceedance at the exact
rank, the number a risk report carries; the dotted line at one is no miss. The
level axis is categorical (four chosen levels), so the lines joining a sample size's
points only guide the eye. A cross below every plotted point on the left, or on the
dotted line on the right, offset and shaded by its $n$, marks a cell where no rank
carries the level at all, $n < 1/\alpha - 1$, since there the shortfall is not one rank.
\S\ref{sec:background} gives the closed form every plotted point can be checked against}
\label{fig:severity}
\end{figure}

\paragraph{Why a one-rank error is not a rounding detail.}
The first objection is that a discrepancy of order $1/n$ at a realistic calibration size
is a curiosity. The answer is arithmetic. A helper one rank short delivers
$(\reqrank-1)/(n+1)$, so the \emph{absolute} coverage miss is $1/(n+1)$, flat in the
level the caller asked for, and small. A risk or alerting report states not coverage but
the exceedance rate $1-\text{coverage}$, and there the same miss is a multiplicative
factor that grows as the level tightens, reaching \muMaxFactor{} times the exact rank's
own exceedance in the sweep of Fig.~\ref{fig:severity}. Both panels are the same
measurement; only the right-hand one depends on what was requested, and it is the one the
report's reader sees. Three further answers, none needed to dispose of the objection,
are the horizon-scaled floor (\S\ref{sec:horizon}), the group-conditional split
(\orsec{D}), and off-grid unattainability at every $n$ (\orsec{G}).

\paragraph{Against the spread of one calibration set.}
The rank a guarantee names fixes coverage in expectation over calibration draws. Any
one caller has a single draw, and its coverage is $\mathrm{Beta}(k, n+1-k)$ at rank $k$. At
$n = 50$ and $0.90$ the one-rank gap is \bsGapFifty{} against a standard deviation of
\bsSdFifty{}, and the caller's own set leaves coverage below the requested level with
probability \bsPBelowFifty{} at the required rank and \bsPBelowShortFifty{} one rank
short; at $n = 1000$ the gap is \bsRatioThousand{} of a standard deviation. Across a
grid of levels and sizes, in exact arithmetic, the gap is under one standard deviation
in \bsCellsGapBelowSd{} of \bsCells{} feasible cells. The deficit is therefore second
order to calibration-draw variability beyond small $n$. It matters to whoever
aggregates: a system that recalibrates often, or a rate averaged over many
calibrations, inherits the shifted mean and none of the spread. A caller who wants a
statement about their own calibration set needs the tolerance-limit rank of
\citet{wilks1941}, a different standard this audit does not score.

\begin{figure}[t]
\centering
\includegraphics[width=0.62\textwidth]{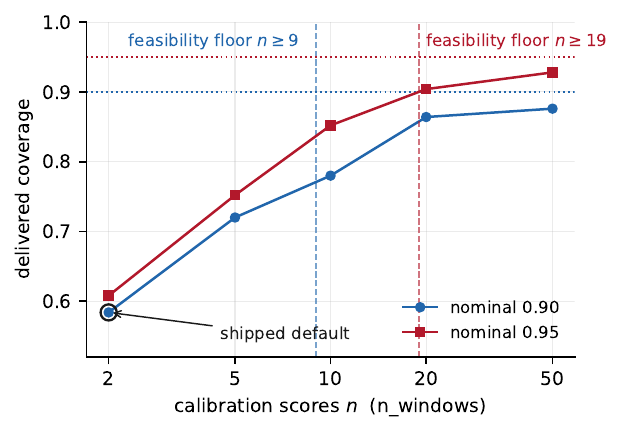}
\caption{The deficit in delivered coverage. Delivered coverage against calibration size for
\code{statsforecast}'s conformal interval, both requested levels, with the feasibility
floor $n \geq 1/\alpha - 1$ marked for each. The shipped default sits on
the wrong side of both, and the circled point is the \sfDefaultNinety{} RQ3 opens
on. \code{n\_windows}, the library's count of calibration windows, is the
calibration size here; \S\ref{sec:realdata} runs the arms behind the figure}
\label{fig:sf}
\end{figure}

This paper audits that conversion in released code: what the libraries a practitioner
installs deliver at the level they are handed, what they return where no rank carries the
request at all, and why the obvious diagnosis of the gap, the presence or absence of the
$(n+1)/n$ correction, is wrong. Words it uses in a narrow sense are collected at the end
of this section.

\paragraph{Research questions.}
The audit asks three questions, each settled by a measurement rather than an argument.

\textbf{RQ1.} Which rule does a shipped resolution site implement, and does reading
its source establish that rule? \S\ref{sec:census} extracts the threshold every site the
suite reaches returns and classifies it against a named candidate set.
\S\ref{sec:bias} answers the second half, no on its own evidence:
\rqOneResisted{} sites resisted classification by reading (named there),
and \rqOneContradicted{} of those \rqOneResisted{} contradicted
what their source suggested, so execution decided the class wherever a site could be
run, and the census marks which. How often a reading \emph{would} suffice is a rate;
\S\ref{sec:conclusions} names the instrument built for it and reports the one sheet
returned so far, not yet a rate.

\textbf{RQ2.} Does the coverage a caller receives match the level requested, and
where it does not, is the $(n+1)/n$ correction what predicts the gap?
\S\ref{sec:realdata} measures delivered coverage against a paired arm built at the
required rank, on two forecasting archives and two tabular suites. The correction
does not predict it. \S\ref{sec:realdata} and \S\ref{sec:severity} give the quantity that does for two of
the three forecasting libraries, and \S\ref{sec:robustness} reports where it fails for
the third.

\textbf{RQ3.} What does a library hand back where no rank carries the requested
level at all? \code{statsforecast}'s \code{ConformalIntervals} defaults to two
calibration windows, \code{n\_windows}$=2$, so $n=2$ calibration scores. A $0.90$
interval there needs rank $\lceil 3 \times 0.90 \rceil = 3$, and only two scores exist.
The library returns a finite interval regardless, and on the first of two forecasting
archives that default delivers \sfDefaultNinety{} against the requested $0.90$
(\S\ref{sec:realdata}). It is the largest shortfall we measure at that level and its
cause is infeasibility, not the conversion RQ1 and RQ2 audit. The best rank available at
$n=2$, the second of two, covers with probability $2/3$; the default delivers less.
\S\ref{sec:boundary} runs every call path we can reach at such a calibration size and
reports what each returns and whether it says anything.
\S\ref{sec:unbounded} follows the path whose own documented keyword removes the
guard.

Beside the three sits a record rather than a question. Every defect this paper reports
was filed upstream before this submission: \dsFilings{} filings over \dsTargets{}
targets, \dsMerged{} of them merged. \deanon{\orsec{I}'s disclosure table carries each
identifier, so a reader can re-query a filing rather than take this paper's word for
it}{Identifiers are withheld here for the reason \S\ref{sec:checks} gives, and the
anonymised manifest accompanying this submission carries the dates and the sites}.
\S\ref{sec:checks} reports, from the trackers rather than from recollection, the state
of each filing on \dsAsOf{} and the reply it drew. It measures response, not agreement,
and is a disclosure record because that is all a window of this length can carry.

\paragraph{Contributions.}
Four findings, in the order the questions are asked; the disclosure record
(\S\ref{sec:checks}) is not a finding.

\begin{enumerate}\itemsep2pt
\item \textbf{A per-helper census} (\S\ref{sec:census}, RQ1): \cenPTwo{} resolution
      sites across \cenLibraries{} packages, each anchored to a file and line and
      classified by execution wherever it can be run, shipped as a conformance suite
      (\S\ref{sec:suite}).
\item \textbf{Delivered coverage on real data} (\S\ref{sec:realdata}, RQ2), against a
      paired arm that differs from the shipped call in the rank alone. The deficit
      concentrates where the level-to-rank map is lossy; the tabular arm is a null for
      undercoverage in all \tabConfigs{} implementations.
\item \textbf{No shared contract at the feasibility floor} (\S\ref{sec:boundary},
      RQ3): a call below it raises, returns an infinite bound, or returns a finite number
      in silence, and \confRouteSilent{} of \confPaths{} paths do the last. A Bonferroni
      split over a horizon $H$ raises the floor to $H/\alpha - 1$ (\S\ref{sec:horizon}).
\item \textbf{A reachable anti-conservative path in a shipped library}
      (\S\ref{sec:unbounded}, RQ3): a documented keyword meant to permit an infinite
      bound skips the guard and delivers \clipDelivered{} against \clipLevel{} at
      $n = \clipN{}$, on \clipReaches{} public regressor classes.
\end{enumerate}

Table~\ref{tab:implications} carries what a caller can do about each, and \orsec{O}
gives each section's claim and the observation that would have sunk it. The proofs
behind \S\ref{sec:imported} are in \S\S A to C of \orone{}. Negative controls that
would have killed the thesis sit in \S\ref{sec:severity}, \S\ref{sec:tabular} and
\S\ref{sec:attributable}: coincidence cells at exact zero by construction, a tabular
implementation exact by construction, and a Ljung--Box screen checked both ways
(\orsec{H}).

\subsection{Terms}
\label{sec:terms}

Two words carry a narrow meaning through the discussion, and both are units rather
than descriptions. A \emph{resolution site} is one expression that turns a requested
level into a position in the calibration scores, or into the denominator of a
$p$-value, and whose value settles what threshold, interval, set or $p$-value comes
back. It is this paper's unit of analysis (\S\ref{sec:census}), and the units nest:
sites sit inside packages and are reached through entry points. The \emph{feasibility
floor} is $n \geq 1/\alpha - 1$, the smallest calibration size at which any rank
carries the requested level. Below it no deterministic rule can deliver that level
(\S\ref{sec:boundary}), and a Bonferroni split over $H$ horizons raises the floor
to $H/\alpha - 1$ (\S\ref{sec:horizon}).

Two more are used in this paper's sense, not the wider literature's.
\emph{Silent} is a property of one instance and never of a branch: a path that
returns a number below the feasibility floor while raising nothing and warning of
nothing, where the harness records exceptions and every Python warning and does not
capture logging output. A site is reported as a \emph{defect} only where what it
returns and what its own documentation or public keyword states disagree; a lossy
convention that is documented is measured and reported, not filed (\S\ref{sec:checks}).
\emph{Delivers} keeps its plain-English sense: whatever coverage a call returns,
predicted or guaranteed. Where that distinction matters it is named,
\emph{predicted} or \emph{guaranteed} coverage.

Further terms are fixed at first use and collected in \orsec{L}, with a figure of the
rank vocabulary on one sorted sample.

\deanon{Who consumes such a threshold decides the right unit of analysis, as two
deployed settings show. In incident management
the requested level is a false-alarm rate, and the cutoff an anomaly detector alerts on
decides whether an operator acts on an alert at all \citep{verma2025observability}. In
donor analytics a cutoff on a predictive score sets who a fundraising programme
contacts \citep{lalakiya2025donor}. Where a library default turns the requested level
into either cutoff, that default is a resolution site, and a system that consumes the
cutoff without reading the quantile call that produced it cannot tell which rank its
level became. That is the argument for counting resolution \emph{sites}
rather than libraries in \S\ref{sec:census}, and for the checklist of
\S\ref{sec:suite} asking which level was \emph{resolved} and not which was
requested: a caller who cannot see the convention cannot report the requested level
and be understood.}{}

\section{Related work}
\label{sec:related}

Each literature is named with what it established and what this paper adds;
\orsec{N} gives the full account.

\paragraph{Defects in numerical and machine-learning libraries.} \citet{tambon2024}
call this class a silent bug: a value returned with no exception and no warning, and
wrong. \S\ref{sec:boundary} finds that pattern between a requested level and the
calibration set that must carry it, and adds that the correct behaviour is not agreed:
\confRouteSilent{} of \confPaths{} public call paths return a finite number below the
floor, \confRouteInf{} return an infinite bound and \confRouteRaise{} raise.
Differential testing against a sibling library \citep{evans2007, pham2019} therefore
reports a disagreement rather than a defect. Bug studies of this ecosystem
\citep{thung2012, zhang2018, islam2019} work from reports and commits; our unit is one
expression, anchored to a file and a line and classified by running it.

\paragraph{API misuse, where the caller is not the one misusing it.} Detectors of API
misuse \citep{amann2019, galappaththi2024} look at callers. Here a caller passes a level
inside the documented range and receives a threshold at the wrong rank, so no rule over
the call site would flag it. The nearest prior art is documentation that contradicts
implementation \citep{zhong2013, xu2025, xie2022, aghajani2019}; \S\ref{sec:unbounded}
is the hard case, a keyword that contradicts its documentation only when a second,
unshown condition holds.

\paragraph{Defaults.} Most configuration errors come from staying with a default
\citep{xu2015}, and options are typically created without being planned
\citep{sayagh2020}. Here the default is a calibration \emph{size}, and whether it clears
the feasibility floor is arithmetic the call does not surface: of the \confPaths{} paths
Table~\ref{tab:conformance} executes below that floor, \confWarns{} warn.

\paragraph{Oracles for numerical code.} With no oracle to hand \citep{weyuker1982,
kanewala2014}, the paired arm of \S\ref{sec:method} is a pseudo-oracle
\citep{davis1981, barr2015} whose relation to its subject is metamorphic
\citep{chen2019, segura2016}: not that either interval is right, but that the two stand
in an order two index rules fix. Library test suites use a hand-chosen tolerance
instead \citep{nejadgholi2019, dutta2020}, and a tolerance wide enough for Monte Carlo
noise absorbs an $O(1/n)$ deficit; \S\ref{sec:certificate} makes the fixture size
decidable so that no tolerance is needed. Audits of released statistical software
\citep{mccullough1998, mccullough1999} score digits against a reference value; here the
oracle is an index, and a threshold at the wrong rank is exact arithmetic on the wrong
order statistic, so no gain in precision moves it.

\paragraph{Conformal prediction and sample quantiles.} The split construction is
\citet{papadopoulos2002}'s, with the theory in \citet{vovk2005} and reviews in
\citet{fontana2023, angelopoulos2023}; the required index is \citet{lei2018}'s and is
stated in TorchCP's library paper \citep{torchcp}. Smoothed $p$-values \citep{vovk2005}
give the valid bound of \orsec{G} where no deterministic one exists. \citet{romano2019}
is the two-rail construction whose conservatism \S\ref{sec:tabular} measures, and
\citet{barber2021} and \citet{tibshirani2019} cover the jackknife+ and covariate shift.
The nine-definition taxonomy is \citet{hyndman1996}'s, reaching this ecosystem through
\code{numpy} \citep{numpy2020}; the interpolated threshold's coverage of a parameter of
$F$ is studied in \citet{hutson1999, beran1993, nagaraja2020}, and \S\ref{sec:imported}
states what exchangeability adds.

\paragraph{Library audits and calibration-set size.} The one peer-reviewed cross-library
comparison we located scores packages on capability and has no correctness dimension
\citep{torchcp}; we supply that dimension. The audited packages are cited by the
reference each pinned distribution names \citep{bostrom2022, mapie2023, loning2019,
herzen2022, montiel2021, mendil2023, garza2022, olivares2022}. \code{quantconformal}
\citep{quantconformal}, first released after this audit's artifact was first archived,
implements what our checklist asks for and, run rather than read, lands on
$\lceil (n+1)(1-\alpha) \rceil$ in all \fsQcCells{} cells we tried. The exact law of
empirical coverage, the minimum calibration size and coverage dispersion are in
\citet{empcov2025, kladny2025, ding2023, gupta2020}; their mechanisms are statistical and
ours is the level-to-rank map, and item~\ref{itm:residue} of \S\ref{sec:imported} shows
that the minimum is the first member of a residue class. \citet{pasche2025extreme} build
an informative interval below the floor from extreme-value statistics.

\section{Background: ranks, levels, and the sample-quantile definitions}
\label{sec:background}

The guarantee is carried by a rank, every shipped interface asks for a level, and the
conversion is where it is lost.

Let $V_1,\dots,V_n$ be calibration scores exchangeable with a test score
$V_{n+1}$, drawn from a jointly continuous law so that ties have probability zero (true of
the synthetic laws here; real residuals can tie, a case the identity does not cover), and let
$V_{(1)} \le \dots \le V_{(n)}$ be the order statistics. Then
\begin{equation}
\Pr\!\left(V_{n+1} \le V_{(r)}\right) = \frac{r}{n+1},
\label{eq:rank}
\end{equation}
where continuity is what makes this an equality: with ties the left side can only
rise, so every bound below is conservative, not wrong, if it fails.
A one-sided bound at rank $r$ therefore delivers exactly $r/(n+1)$, and the smallest
rank achieving a requested $1-\alpha$ is
$\reqrank = \lceil (n+1)(1-\alpha) \rceil$. When $\reqrank > n$ no finite
\emph{deterministic} bound built from these scores is valid, which happens precisely
when $n < 1/\alpha - 1$; \orsec{G}
exhibits the randomised bound that is valid there; at $\alpha = 0.10$ the floor is
$n \ge \rmTheoreticalNinety{}$ and at $\alpha = 0.05$ it is
$n \ge \rmTheoreticalNinetyFive{}$. Both facts are standard
\citep{angelopoulos2023, ding2023}; every measurement below is expressed against them.

The multiple Fig.~\ref{fig:severity} plots has a closed form. A helper landing one rank
low leaves $(n+2-\reqrank)/(n+1-\reqrank)$ times the exceedance the exact rank leaves, so
every plotted point can be checked by hand: $6/5$ at $n=50$ and $0.90$, $3/2$ at $n=50$
and $0.95$ and again at $n=200$ and $0.99$, and $2/1$ at $n=\muWorstN{}$ and
$\muWorstLevel{}$, the worst cell in the sweep. Exact arithmetic.

The gap this paper is about opens when $\reqrank$ must be requested through a
level. \citet{hyndman1996} catalogue nine definitions of the sample quantile,
three based on rounding and six on interpolation, and their taxonomy is
implemented directly by the \code{method=} argument of \code{numpy} \clVerNumpy{},
the release every count below is read off, which exposes four
further methods beside the nine, kept for backward compatibility rather than drawn from
that taxonomy: \code{lower}, \code{higher}, \code{midpoint} and \code{nearest}. A caller therefore chooses among \rmDefs{} names, and \rmDefs{}
is the denominator every count below uses; nine is Hyndman and Fan's taxonomy and
\rmDefs{} is what ships. Hyndman and Fan
compared the definitions as \emph{estimators of a population quantile}. We
compare them as \emph{carriers of a coverage guarantee}, which is a different
question with a different answer: an estimator may be excellent and still be
unable to express $\reqrank$ at any level.

Write $h$ for the \emph{virtual index} a definition computes from the level: the
position, generally not an integer, that it interpolates around. For the six
continuous definitions $h$ is affine in the level, $h = A + Bq$ with $A$ and $B$
fixed by the definition and $n$; for five of the seven rounding definitions $h$ is an
integer; \code{midpoint}, which averages two adjacent order statistics, is a
half-integer by construction; and \code{averaged\_inverted\_cdf} has
$h = qn + \tfrac12$ when $qn \in \mathbb{Z}$ and $h = \lceil qn \rceil$ otherwise.
The pairs are not ours: $A$ and $B$ are \citet{hyndman1996}'s
$(a_{\mathrm{HF}}, b_{\mathrm{HF}})$, with $A = a_{\mathrm{HF}}$ and
$B = n+1-a_{\mathrm{HF}}-b_{\mathrm{HF}}$. Hyndman and Fan give these as $\alpha$ and
$\beta$; those letters are taken here, so the subscript marks whose they are, inherited in turn
from the plotting positions of \citet{blom1958} and \citet{cunnane1978}, surveyed by
\citet{harter1984}. We read $h$ off the shipped implementation rather than its
documentation, by quantiling the tie-free score set $1,\dots,n$, whose values
\emph{are} their own ranks, and check it against the published
$(a_{\mathrm{HF}}, b_{\mathrm{HF}})$ pair.

\begin{proposition}[The guarantee of an interpolated threshold]
\label{prop:interp}
For a definition that returns $T = (1-\gamma)V_{(j)} + \gamma V_{(j+1)}$ with
$\gamma \in (0,1)$, the threshold lies between two order statistics, so
\eqref{eq:rank} does not apply to it: the distribution-free guarantee it carries
is $j/(n+1)$, that of rank $j = \lfloor h \rfloor$, and the coverage it attains
satisfies $j/(n+1) \le \Pr(V_{n+1} \le T) \le (j+1)/(n+1)$, the upper inequality
being strict whenever $\Pr(T < V_{n+1} \le V_{(j+1)}) > 0$.
\end{proposition}

\begin{proof}
Since $\gamma \in (0,1)$ and $V_{(j)} \le V_{(j+1)}$, the convex combination $T$
lies in $[V_{(j)}, V_{(j+1)}]$. Hence
$\{V_{n+1} \le V_{(j)}\} \subseteq \{V_{n+1} \le T\} \subseteq
 \{V_{n+1} \le V_{(j+1)}\}$, and \eqref{eq:rank} at ranks $j$ and $j+1$ gives the
two bounds. The upper one is an equality exactly when $V_{n+1}$ never lands in
$(T, V_{(j+1)}]$, so it is strict under any law with mass there but not
distribution-free. Only the lower bound holds for every exchangeable law, which is
what makes it the guarantee. \S\ref{sec:realdata} and \S\ref{sec:horizon} report
the floor it establishes beside every corrected level.
\end{proof}

Proposition~\ref{prop:interp} is the worst case, which is what a
\emph{guarantee} means. A practitioner measures instead the fractional rank, the
interior prediction it supports and the level that inverts them, all three imported
rather than proved here and stated in \S\ref{sec:imported}.

\subsection{Two objections, answered here}
\label{sec:objections}

\paragraph{The obvious diagnosis, and why it fails.}
The natural hypothesis is that libraries which omit the $(n+1)/n$ correction
undercover and libraries which apply it do not. We find that this does not
predict severity. \code{darts} and \code{sktime} both omit the correction and sit
at opposite ends in whether the required rank is ever reached: \code{darts} passes its
uncorrected level through a rounding-based quantile method that lands on an integer
rank while \code{sktime} passes two uncorrected levels through linear interpolation
on signed residuals and reaches the required rank at no calibration size. Conversely, a library can apply
the correction and still return a threshold at a calibration size where no valid
\emph{deterministic} bound exists at all and, more usefully, three libraries
reach the required rank by design, two of them by never passing a level to a quantile
function.

\paragraph{``So pass a different string.''} This objection fails three ways.
\textbf{(i)} Two corrected levels appear in the literature and in shipped code, and the substitution turns on
which one is passed. At $(1-\alpha)(n+1)/n$, \code{method='higher'} lands on the
required rank at \rcHigherExactNinety{} of the \rcHigherSizesNinety{} sizes that
sweep covers at nominal $0.90$ and one rank past it on a residue class of density
exactly $\alpha$;
\code{inverted\_cdf} resolves that level at every $n$ by identity in exact arithmetic
(executed in doubles it lands one past at \rcExecMiss{} of \rcExecCells{} cells, all where
$(1-\alpha)(n+1)$ is an integer and the corrected level is not a double; \orsec{C}), it is the only call
returning an actual order statistic to do so at all \rcLevels{} levels checked, and no
tutorial gives it. At
$\lceil (1-\alpha)(n+1) \rceil/n$ the string almost stops mattering:
\rcCeilExactNinety{} of the \rcCeilDefs{} definitions return the required rank at
every one of the \rcCeilSizesNinety{} feasible sizes at $0.90$, a separate sweep on
its own grid, and \code{higher} lands one rank wide at \rcCeilHigherOverNinety{} of
them. What repairs the call
is therefore the level, not the string. The level is still a lossy channel, in the one
direction that matters. $\reqrank/n$ is an integer over $n$ and no double holds it
exactly at \rcCeilInexactNinety{} of those sizes; at \rcCeilShortNinety{} of them
\code{interpolated\_inverted\_cdf} floors a position exact rationals put at $\reqrank$
and hands back $\reqrank - 1$. \orsec{C} gives that arithmetic, the two exceptions, and
why the canonical idiom, \code{higher} at $(1-\alpha)(n+1)/n$, is not the best fix. \textbf{(ii)} The deficit does not shrink with
data. The delivering set is periodic in $n$, so a call honest at one calibration size is
short at the next, and at an uncorrected level only \rmDeliverNinety{} of \rmDefs{}
definitions ever guarantee nine tenths at all (item~\ref{itm:residue} of \S\ref{sec:imported}). \textbf{(iii)} The
level-to-rank map is one defect of several:
\S\ref{sec:boundary} is a feasibility floor with no shared contract, \S\ref{sec:align} an
index error in the score set rather than in the level, \S\ref{sec:unbounded} a documented
keyword doing the opposite of its documentation, and \S\ref{sec:realdata}'s largest cell a
shipped default below its own floor. No choice of interpolation string repairs any of them.

\subsection{Three results the audit rests on}
\label{sec:imported}

The coverage numbers in this paper are read against three results about the
level-to-rank map. None of them is a property of a conformal-prediction library: each
is a property of the sample-quantile definitions and of the interfaces that expose
them. Each is proved in \orone{}, \S A to \S C, and stated here precisely enough to use
and to falsify. The span arithmetic the extension of
item~\ref{itm:cert} to all $n$ rests on is also re-checked in exact rationals over every
$n \le 2000$ before a series is read.

The paper proposes no new conformal method. The required index, the feasibility
condition $n \geq 1/\alpha - 1$ and the Beta law for an order statistic's coverage are
established \citep{angelopoulos2023, ding2023}. The distribution-free fractional-rank
identity, and the tail law that gives $\pi$ in closed form through the extreme-value
shape $\xi$ and makes the gap \emph{depth} rather than the sample size the governing
variable, are proved in \orsec{A} and \orsec{B}, and a Monte Carlo sweep matches the
closed form to \glWorstLaw{} over \glCells{} cells, and to \glOffWorst{} over
\glOffCells{} more whose sizes are not divisible by ten.

\begin{enumerate}\itemsep3pt
\item \label{itm:frank}\textbf{Fractional rank.} A threshold interpolated at fraction $\gamma$ between
      $V_{(j)}$ and $V_{(j+1)}$ delivers $(j+\pi)/(n+1)$, where $\pi$ is the
      conditional position of the test point in the gap, and that identity is
      distribution-free. Under i.i.d.\ sampling $\pi = \gamma$ when the density is
      flat across the gap, so predicted coverage is $h/(n+1)$ for the virtual index
      $h$ in the interior; \orsec{A} gives the closed form for $\pi$ that the tail
      law of \orsec{B} builds on, and the hypothesis it carries. Every simulated cell is i.i.d.\ by
      construction; real series are not, and \S\ref{sec:realdata} says what that costs. \emph{Used in:} \S\ref{sec:realdata}, \S\ref{sec:horizon},
      \orsec{D}.
\item \label{itm:residue}\textbf{Only \rmDeliverNinety{} of \rmDefs{} definitions ever guarantee a
      requested nine tenths, and they guarantee it on a residue class in $n$}, not above a
      threshold, so $n_{\min}$ is the smallest member of a periodic set and
      ``collect more data'' is not the remedy. At a level $p/d$ in lowest terms with
      $\alpha < \tfrac12$, the density of that set is $\alpha$ for \code{inverted\_cdf} and
      \code{averaged\_inverted\_cdf}, $1/d$ for \code{weibull}, $2\alpha$ for
      \code{higher}, and zero for every other continuous definition; \orsec{C} proves all
      five. The rounding definitions that never deliver are read off a sweep to
      $n = 2000$ at the levels swept, not proved. The same residues decide where the
      canonical idiom sits. \emph{Used in:} \S\ref{sec:background}, \S\ref{sec:severity}, where one
      library's coincidence cells turn out to be exactly this set, and
      \S\ref{sec:checks}.
\item \label{itm:cert}\textbf{A minimal certificate.} The \ceRules{} candidate resolution rules fall
      into \ceClasses{} observationally distinct classes, separated by \ceCertSize{}
      calibration sizes at each of the \ceLevels{} levels checked. The count came
      out the same at all three and the sizes did not, which is an observation about
      three levels and not a theorem. No single size separates them. Above a
      computable threshold every candidate's delivered minus required rank is periodic
      in $n$, so the irregular prefix checked exhaustively plus two full periods covers
      every $n$; \orsec{C} proves the periodicity. The certificate is relative to its
      candidates and says nothing about a rule outside them. \emph{Used in:} \S\ref{sec:suite} and
      \S\ref{sec:checks}.\end{enumerate}

The audit's own numbers depend on two consequences of item~\ref{itm:frank}. The distribution-free \emph{guarantee} of an interpolated threshold is
$\lfloor h \rfloor/(n+1)$, so a helper reporting ``rank $r$'' in the sense of the
smallest order statistic at or above its threshold guarantees $r-1$, not $r$. And
predicted coverage being $h/(n+1)$ rather than $\reqrank/(n+1)$ is why the paired
design of \S\ref{sec:realdata} compares a library against \emph{its own} scores at
the required rank, and not against a nominal level.

\paragraph{The fractional rank, the interior prediction, and the level that inverts
them.} Three expressions carry the rest of the paper: what an interpolated threshold
actually delivers, what that becomes when the density is well behaved, and which level a
caller has to pass to get the rank they wanted.

What recovers the interpolation Proposition~\ref{prop:interp} discards is the
\emph{fractional rank}: writing $\pi$ for the conditional
position of the test point within the gap,
$\pi = \Pr\big(V_{n+1} \le T \mid V_{(j)} < V_{n+1} < V_{(j+1)}\big)$,
\begin{equation}
\Pr\!\left(V_{n+1} \le T\right) = \frac{j + \pi}{n+1}
\label{eq:frac}
\end{equation}
for every exchangeable law, with $\pi = \gamma$ whenever that conditional law is
uniform on the gap and $\pi = (n{+}1)\,\mathbb{E}[F(T)-F(V_{(j)})]$ under i.i.d.\
sampling. The first proposition of \orsec{A} is the statement
with its proof and with the counterexample separating the two hypotheses.

The closed form is a \emph{ratio of expectations}, the denominator being $1/(n+1)$, and taking
it as the definition invites the expectation of a ratio instead, which can match a
simulation to four decimals while contradicting \eqref{eq:frac}.

Three things follow, and they are what separate this paper from a rank-arithmetic
audit. First, Proposition~\ref{prop:interp} is the $\pi = 0$ corner of
\eqref{eq:frac}; the deficits we measure are one rank because $\pi$ is bounded by
$1$, not because interpolation is discarded. Second, where the density is locally
Lipschitz \emph{and bounded away from zero} the gap is $O(1/n)$ wide and
$\pi \to \gamma$ there: positivity is what buys the gap width, and where the
density vanishes at the quantile the rate below is lost. Where it does not,
\begin{equation}
\Pr\!\left(V_{n+1} \le T\right) \;=\; \frac{h}{n+1} + O(1/n^2)
\label{eq:hpred}
\end{equation}
with $h = j + \gamma$ the virtual index and the interior rate, at depth $n - j$ of
order $n$, \citet{beran1993}'s, one expression covering all
\rmDefs{} definitions in the interior, which reduces to \eqref{eq:rank} exactly when $h$ is an
integer. Third, \eqref{eq:hpred} is invertible. Solving $h(q) = (1-\alpha)(n+1)$
gives the level at which a definition delivers what was asked:
\begin{equation}
q^{\dagger} = \frac{(1-\alpha)(n+1) - A}{B},
\label{eq:qdag}
\end{equation}
which for \code{numpy}'s default \code{linear} ($A=1$, $B=n-1$) is
$\big((1-\alpha)(n+1) - 1\big)/(n-1)$, and \emph{not} the corrected level in general
use. That level has two published forms:
$(1-\alpha)(n+1)/n$, which is \citet{romano2019}'s, and
$\lceil (1-\alpha)(n+1) \rceil / n$ with the ceiling taken before the division, which
is \citet[p.~\pfBodyPage{}]{angelopoulos2023arxiv}'s. \citet{lei2018} state the requirement as a rank and
form no level, and so does Romano's released code, which selects the
$\lceil (1-\alpha)(n+1)\rceil$-th score directly. A third spelling, in the appendix
code of the same tutorial \citep[p.~\pfAppendixPage{}]{angelopoulos2023arxiv}, takes the
ceiling over the whole ratio and so resolves to $1$, the sample maximum, at every
feasible size. Which rank the first lands on depends on the definition:
\code{inverted\_cdf} reaches the required rank at every $n$, by an identity,
\code{higher} lands on it except on a residue class of density $\alpha$, where it
lands one rank past, and every interpolating definition lands short of that rank while overshooting the level it was handed.
Item~\ref{itm:residue} above gives that
class in closed form. The second
is the required rank written as a fraction of $n$, so a definition that rounds
recovers the rank, or with \code{higher} lands one past it; \S\ref{sec:objections} reports what the \rmDefs{} definitions return
when handed it.

\eqref{eq:qdag} buys asymptotic exactness, not a finite-sample guarantee: the
distribution-free floor of a threshold at virtual index $h$ is still
$\lfloor h \rfloor/(n+1)$ by Proposition~\ref{prop:interp}. A reader who needs the guarantee should use a rounding definition at the
required rank. A reader who wants the level to mean what it says, as predicted
coverage, can pass it to \code{weibull}, whose virtual index $q(n{+}1)$ (for $q \le n/(n{+}1)$) makes the prediction
\eqref{eq:hpred} equal to $q$ in the interior; one
who cannot change the definition can use $q^{\dagger}$ and accept an $O(1/n^2)$ interior error and the tail regime of
\orsec{B}.

\section{Study design}
\label{sec:method}

Reading a helper's source only guesses what it will do; \S\ref{sec:census} reports
this paper guessing wrong on a package of its own choosing, where reading missed a site
and running it did not. This section replaces the guess with an execution, and pairs
the arms so that a delivered coverage is attributable to the index alone.

\subsection{Subjects and data}
\label{sec:subjects}

The subjects are \cenLibraries{} Python packages that resolve a distribution-free
coverage bound from a calibration set, at the pinned versions of
Table~\ref{tab:census}. How the candidate list was assembled, which packages it
excludes and on what criterion, and what share of the ecosystem's downloads the
audited set carries are in \S\ref{sec:census}, with the recall of two searches against the named list
measured in \orsec{F}.

The forecasting arms run on the Monash Time Series Forecasting Archive
\citep{godahewa2021}, loaded through \code{sktime}'s \code{load\_forecastingdata}.
The primary collection is M1 monthly \citep{makridakis1982} and the replication of
\S\ref{sec:replication} is M3 monthly \citep{makridakis2000}, a separate
collection. The headline cells take the first \skSeries{} series in archive order above
a length floor, six observations beyond the largest calibration setting the arm runs
for \code{sktime} and \code{statsforecast} and the longer floor its series cache
needs for \code{darts}, which \orsec{H}'s whole-archive floor is built on, and give
each one test point; \S\ref{sec:robustness} removes the
first of those two choices by rerunning every cell on all \robSeries{} eligible
series of the \robArchive{} in M3 monthly at
\robOrigins{} rolling origins each. The frequency-and-model arm adds M1 and M3 quarterly and M4 daily
\citep{makridakis2020}, and caps the daily series at a fixed length so the boosted
forecaster refits an affordable number of times; the probe output states the cap
and the number of series it reached.

The tabular arms run on two curated OpenML suites \citep{vanschoren2014}:
OpenML-CC18 for classification \citep{bischl2021} and OpenML-CTR23 for regression
\citep{fischer2023}. The selection rule was fixed before any measurement and is
the same for both: at most \tabMaxRows{} rows, at most \tabMaxFeatures{} features
after one-hot encoding, no missing values, taken in dataset-id order, and the first
\tabDatasetsClf{} that pass. The cap binds on classification only; regression runs out
of qualifying datasets under it. That keeps
\tabDatasetsClf{} classification datasets and \tabDatasetsReg{} regression
datasets; \tabSkipClf{} and \tabSkipReg{} are skipped, and the probe output names
every one of them with the clause it failed. Each cell draws a fresh random
permutation of a dataset's rows from a seed fixed per calibration size: half of the
rows left after the calibration set, and at least thirty, train the base model, the next $n_{\text{cal}}$
calibrate it, and the remainder, capped at a thousand rows, is the test set, all shared
by both arms.

\subsection{Classifying a resolution site}
\label{sec:classify}

\paragraph{Threshold extraction, not set comparison.}
Two behaviours are indistinguishable from returned intervals alone: clamping an
out-of-range level to $1.0$, and never correcting the level at all. Both can
return the maximum calibration score. We therefore extract the returned
\emph{threshold} and identify which order statistic it corresponds to, using a
tie-free score set $1,\dots,n$ so that a returned value \emph{is} its rank.

\paragraph{Two probes, because one is not enough.}
A threshold equal to $\max(\text{scores})$ is either a clamped level \emph{or} the
genuine rank-$n$ order statistic (indistinguishable where $\reqrank = n$), and
at some $n$ an uncorrected rule coincides with a corrected one, a confusion that
cost us a wrong classification before we caught it. The classifier therefore (i)
fits the returned ranks over six interior $n$, namely $n_0$, $n_0{+}1$, $n_0{+}3$,
$n_0{+}7$, $2n_0$ and $3n_0{+}1$ with $n_0 = \max(n_{\mathrm{f}}{+}20,\, 40)$ for $n_{\mathrm{f}}$
the feasibility floor of \S\ref{sec:terms} at the level under test, not the
$n_{\min}$ of \S\ref{sec:suite},
against a named candidate set:
the cross product of rail convention (one-sided or two-sided), correction (raw,
or clipped at $1$), and family (\code{linear}, \code{higher},
\code{inverted\_cdf}, or a direct order statistic), and (ii) probes an $n$ below the feasibility floor.

\paragraph{Why we do not fit curves.}
A helper matching no candidate reports an incomplete candidate set, not an
approximate library. Three times during this work the difference mattered: a
two-sided construction read as an approximate estimator, a coincidental $\max$
read as a clamp, and a returned \emph{count} read as a threshold, each a
confident wrong classification, each now a regression test that fails at import.

\paragraph{Exact arithmetic and self-checks.}
Every closed form validates itself against exact rational arithmetic at import,
on a grid chosen by the boundary rather than by round numbers, and a failing
check aborts the run. Sweeps extend to at least twice the first boundary found
(stopping just past one looks converged and is worse than stopping far short), and every sweep
includes a non-unit-fraction level, since the residue
structure of \S\ref{sec:severity} collapses to a clean pattern only for unit
fractions.

\paragraph{The branches a helper can be in.}
\label{sec:taxonomy}
At the floor, \code{numpy.quantile} raises for a level above $1.0$, so an
implementation that corrects the level must do \emph{something} there. The
distinct behaviours we observe are: (a) let the error propagate; (b) clamp the
level to $1.0$, returning the maximum score where the conformal quantile does not
exist; (c) return $\pm\infty$, which is vacuous but honest; (d) apply no
correction at all, so the floor is never reached and no choice is ever made;
(e) non-smoothed conformal $p$-values, vacuous exactly when
$1/(n+1) > \alpha$ under the inclusion rule $p > \alpha$; (f) smoothed $p$-values, exactly uniform when the tie term is
the published one and therefore exact at every $n$, bought with randomisation
(\orsec{M} names the shipped expression at a pinned version that carried a
different tie term, which we filed and which merged); and
(g) an approximate streaming quantile estimator with no correction and no warm-up
guard.
Silence is a per-instance attribute, not part of any branch's definition: every
branch (c), $\pm\infty$, path we execute below the floor warns (\S\ref{sec:boundary}),
but nothing in the branch requires it.
The unit of analysis is the resolution site, not the library: a
single package can resolve the bound several different ways in several code paths,
and any table with one row per library is structurally unable to say so.

\subsection{The two arms}
\label{sec:arms}

\paragraph{Two arms, and what changes between them.}
Every coverage measurement below is a paired difference. \textbf{Arm A} is the
library through its public API at its own nominal level. The probes choose the
base forecaster (\S\ref{sec:realdata}) and set one argument of the conformal helper,
its calibration size (\code{sktime}'s initial window, \code{statsforecast}'s window
count, \code{darts}' calibration length), which Table~\ref{tab:forecasting}'s
setting column names, and leave every other argument of the helper as shipped; the
subject is what ships, not what is achievable.

\paragraph{The standard, and whose it is.}
The forecasting helpers measured here document an empirical quantile of their
residuals, not a finite-sample guarantee, and do not promise the required rank. The
required rank is the standard their own conformal naming and the methods they cite
invite, and it is the one a caller who asked for a conformal interval is relying on.
Where a library documents an empirical quantile, a gap to that rank is reported as a
departure from that standard and not as a broken promise; only the tabular helpers,
whose documentation does claim the guarantee, are held to it as a contract, and there
the result is a null (\S\ref{sec:tabular}). \textbf{Arm B} takes \emph{the
library's own calibration scores, the library's own point prediction and the
library's own interval geometry}, and changes exactly one thing: the level becomes
the required order statistic. What that is depends on the geometry. For a symmetric band on absolute scores it is the rank
$\reqrank$, or $+\infty$ when $\reqrank > n$. For a helper resolving \emph{two}
levels on signed scores it is the index pair $(a, a+\reqrank)$ spanning $\reqrank$ of
the $n+1$ gaps, since $\Pr(V_{(a)} < V_{n+1} \le V_{(b)}) = (b-a)/(n+1)$ makes a \emph{span} the object such a helper has to deliver (\orsec{K}); the
$n+1-\reqrank$ excluded gaps are split as evenly as the arithmetic allows, and a rail
goes to $\pm\infty$ where there are fewer than two to split. That last case is the
two-rail feasibility floor, $n \ge 2/\alpha - 1$, which is \S\ref{sec:horizon}'s
statement that the floor scales with whatever divides $\alpha$ at $H = 2$.

The geometry clause is summarised in \S\ref{sec:checks}, and \orsec{I} records
what failed, what caught it, and the generalisable lesson.

\paragraph{The index figure, which is two different figures.}
For a symmetric band, ``lands on rank $r$'' means the smallest order statistic at
or above the returned threshold: the rank if the helper returns an actual order
statistic, and $r-1$ by Proposition~\ref{prop:interp} if it interpolates. The
rank the rule \emph{resolved} is the index its level-to-rank map names; the
landed rank is what the returned threshold selects in the realised score set,
for the prediction column recovered from the interval half-width rather than
read off the quantile call, and it is the figure that column averages. A
two-rail helper has no such $r$: half of an asymmetric width is not an order
statistic of anything, and the helper returns two thresholds rather than one, so
what we report is the \emph{span}
$\lfloor h_{\text{hi}} \rfloor - \lceil h_{\text{lo}} \rceil$, where $h_{\text{lo}}$
and $h_{\text{hi}}$ are the virtual indices the lower and upper rails resolve to, that
Proposition~\ref{prop:interp} guarantees across the pair, both rails'
interpolation already discarded. Reporting a rank for a two-rail helper
understates its deficit and names the wrong mechanism.

\subsection{Estimands, uncertainty, and what is tested}
\label{sec:estimands}

\paragraph{The unit, per arm.}
In the archive arms one series is one unit and contributes one test point, so a
cell is \skSeries{} independent binary outcomes per arm. In
\S\ref{sec:robustness} the unit is still the series and the observations are
\robOrigins{} rolling origins within it, which share almost all of their history:
every standard error there is clustered by series and the naive column beside it is
printed only so the dependence is visible. In the tabular arms one dataset is one
unit. Each dataset's own held-out rows give its coverage, and the paired difference
is averaged over datasets rather than over rows, so a single wide dataset carries no
more weight than a narrow one. In the synthetic
arms one series is one unit with one test point, so the errors are binomial and
nothing needs clustering.

\paragraph{Two estimands, and only one of them can fail.}
Every cell reports delivered coverage per arm and the paired difference $\Delta$
between them.

In every forecasting arm a geometry-matched arm B contains arm A, so
$\Delta \ge 0$ pathwise and every unit that changes status there changes the same way;
in \pairedLossCells{} cells across \pairedLossConfigs{} tabular configurations the
containment runs the other way
(below). The containment is a metamorphic relation
\citep{chen2019, segura2016}: two executions sharing every input but the index, whose
outputs must stand in a fixed order whatever the data, asserted at every fit. Neither
execution has an oracle, and the relation does not need one. That is a property of the
two index rules, not of the data, so
$\Delta > 0$ is not a finding, and a standard error beside it is not testing a null
the construction permits. A two-sided test of $\Delta = 0$ tests a boundary of the
parameter space, attainable only if no test point ever falls between the two
intervals, and we do not report one. The standard errors are the sampling error of
the measurement and nothing else.\footnote{No rank-based effect size is reported, for the same reason. The two arms are evaluated on the same units and arm~B contains arm~A, so the outcome pairs are nested indicators rather than two independent samples, and a Mann--Whitney or Vargha--Delaney statistic measures a stochastic dominance the construction already fixes: for two binary outcomes $\hat{A}_{12}$ is $\tfrac{1}{2} + \Delta/2$, a relabelling of the difference the tables already print. The interpretable effect size is $\Delta$ in coverage units, with its multiplicative form in Fig.~\ref{fig:severity} and its rank form $(n+1)\Delta$, the deficit in order statistics.}

The comparison that \emph{can} fail is $\Delta$ against the value the index rules
predict for it, $(\reqrank - h)/(n+1)$ for a symmetric arm, with $h$ the fractional
position arm~A's returned threshold lands at among the $n$ scores, and the span
difference over $n+1$ for a two-rail one, carried in
Table~\ref{tab:forecasting} and \orsec{H}'s robustness table as the per-cell mean of
$(\reqrank_i - h_i)/(n_i+1)$ over feasible units and never a ratio of
medians. A cell agrees with its prediction when the two lie within two standard errors
of each other, clustered by series in \S\ref{sec:robustness}, and fails the comparison
beyond that. The figure it uses is the position arm~A \emph{lands on}, not the rank its
quantile rule \emph{resolved}: the required rank sits at or above the resolved
rank for a symmetric arm, and for a two-rail arm the span rule puts $a$ at or
below $\lfloor h_{\text{lo}} \rfloor$ and $a + \reqrank$ at or above
$\lceil h_{\text{hi}} \rceil$, checked in exact arithmetic over every
$n \le 2000$ before a series is read and asserted on the returned values at every
fit. Over the \pairedCells{} paired cells (\pairedSkCells{} \code{sktime}, \pairedSfCells{}
\code{statsforecast} and \pairedDaCells{} \code{darts} cells across both archives and both
levels, plus \pairedTabCells{} in \S\ref{sec:tabular}'s tabular suite),
\pairedUnits{} unit-level comparisons in all, \pairedReversals{} units lost
coverage under arm B (a series by a test point arm A covered and arm B did not, a
tabular dataset by a lower covered proportion under arm B), concentrated in \pairedLossCells{} cells across
\pairedLossConfigs{} of \S\ref{sec:tabular}'s implementations, the asymmetric
configuration among them, where arm A over-covers by construction. The tables print a
subset of these cells; every one of them, with its discordant counts, is in the
committed probe outputs.

Each arm prints its discordant counts as integers, and the intervals on $\Delta$ are
Wilson score intervals on that count. \orsec{H} gives them, reconciles them with the
discordant-count check of \S\ref{sec:limitations}, and states what a cell is, how
many replications each arm carries and what reproduces. How a language model was used
is stated under \emph{Declarations}.

\section{The empirical map}
\label{sec:census}

Which rule does each shipped site implement? The map below answers site by site, each
anchored to a file and a line: \cenDriven{} of the \cenPTwo{} classified by the
threshold they return when run and \cenShared{} by the threshold returned by the
\code{numpy} method they share. Table~\ref{tab:census} counts them by package and
\orsec{J} lists every one with its anchor, its branch and whether the conformance suite
reaches it.

We downloaded and read \cenRead{} packages at pinned versions. Counting over source
requires a criterion, so we state it and report what it excludes.
\cenExcluded{} are excluded and stay excluded. \code{uq360} takes a percentile across
predictions made under perturbed parameter draws. That is a percentile of a predictive
distribution and not an order statistic of anything held back, so nothing about
finite-sample coverage follows from where it lands. \code{venn-abers} does turn a level into a
whole-number position on a held-out split, but it sorts the calibration
response labels, and those order statistics trim the isotonic calibrators' input rather
than becoming an endpoint, so no miscoverage level is resolved. Both
exclusions are anchored to a file and a line in the artifact's
\code{PLAN.md} and re-opened on disk by the frame probe every time it runs, so the
criterion is applied to each rather than asserted over it.

\cenCorrected{} was excluded and should not have been, and the audit's own instrument
said so. We set \code{conformal-tights} aside after reading
\code{\_weighted\_quantile}, whose quantiles are gradient-boosted model outputs and rank
nothing held back. That was true of the function and false of
the package: its conformal block resolves a requested level over conformity scores on a
held-out third split, which is a resolution site under the criterion above. Reading missed it and running it did not. The out-of-census arm below reports what it delivers, because the package sits
in a separate pinned environment and the census's counts are fixed at the versions it
names. It is the failure \S\ref{sec:bias} reports at the \rqOneResisted{} sites that
resisted reading, occurring inside the census's own sampling frame, and it is why the headline
criterion is applied by execution wherever a reading is contestable.

The remaining \cenLibraries{} resolve a coverage bound from a calibration set and are
the subject of the audit.

\paragraph{The sampling frame.}
The candidate list was assembled by name rather than by a query over the package
index, and the artifact's \code{PLAN.md} enumerates it in full: the \cenRead{}
packages read here, one row each with its verdict and its anchor, and separately the
\iwUnaudited{} compared against them in the download denominator below as queried on
\iwDate{}. The \cenRead{} are the \cenLibraries{} of the census and the
\cenExcluded{} and \cenCorrected{} set aside above; \orsec{F} names the \iwUnaudited{}. The
two populations overlap in \code{venn-abers} and
\code{conformal-tights}, and neither contains the other. Each criterion
is stated ahead of the package it excludes and each exclusion names the criterion it
fails, so a reader who rejects a criterion can see which package returns to the set.
A listed name that resolves to no distribution on the index is recorded as
such, not as a gap. The frame is not a systematic search;
\S\ref{sec:limitations} carries the consequence.

\paragraph{Ten packages, and the denominator.}
On \iwDate{}, over the \iwAudited{} audited packages and \iwUnaudited{} conformal,
quantile-regression and prediction-interval packages we did not audit, the audited set
drew \iwShare{}\% of the two sets' combined monthly PyPI downloads, \iwAudTotal{}
against the unaudited \iwUnaudTotal{}, rising to \iwCfShare{}\% among packages
whose primary purpose \emph{is} conformal prediction, \iwCfAudTotal{} against
\iwCfUnaudTotal{}. The first answers whether these
packages matter, the second whether we audited the ecosystem or a sample of it.
Downloads bound attention, not adoption: a CI job pulling a wheel counts every time and a
vendored package counts once. They say nothing about R, Julia, SQL or spreadsheet callers.

\paragraph{Where the taxonomy stops, checked, not assumed.}
The audited set is tabular and forecasting conformal prediction. A fair question is
whether the level-to-rank map reaches the conformal methods now embedded in
language-model pipelines, and through agent frameworks that wire one shared
module to every agent calling it, so a convention set once reaches callers who
never see it\deanon{ \citep{venugopal2025agents}}{}. We checked one such module.
\code{torchcp}'s \code{llm} package, which implements conformal language modelling,
contains no resolution site under our criterion: no requested level is ever turned
into an index into a calibration set. Its calibration selects a configuration by
fixed-sequence binomial testing over candidate thresholds: Learn-then-Test risk
control, whose guarantee comes from a concentration bound rather than an order
statistic. None of its three \code{torch.quantile} calls is passed $\alpha$, the property that matters. The core helper we do
audit is not reachable from any \code{llm} code path. Risk control by concentration
and rank-based distribution-free coverage are sibling families, and the arithmetic
here applies to the second.

\paragraph{The per-helper count.}
Within those \cenLibraries{} we define a \emph{resolution site}: one expression that
turns a requested level into either an index into the calibration scores or a
$p$-value denominator, and whose value determines a returned threshold, interval,
set or $p$-value. Two public methods sharing one expression count once; one method
containing two different rules counts twice. Under that criterion there are
\cenPTwo{} sites (Table~\ref{tab:census}), reached through \cenPThree{} public API
entry points; the two counts need not agree, because some entry points route to
several sites and some sites are reached from several entry points. Every site is
reachable from at least one, which the census asserts, not assumes. The count is a
lower bound under its criterion, not a proof that nothing was missed: a second read
by a language model found \scAddedAfter{} further sites, \code{mapie}'s
APS cross-validation count and two weighted \code{torchcp} predictors, each confirmed
against the pinned source before the census took it in, and the two \code{torchcp}
sites were also run (\orsec{J}). The
\cenPTwo{} split by disposition, which \S\ref{sec:boundary} sets out and closes to the
\confPaths{} call paths of the conformance table.

The census maps the code, not the issue trackers, and makes no claim
about which of these sites was known to whom. Among them are a second uncorrected
helper in one package that is reachable from a public forecaster, an
exact-by-construction helper on a base class that every calibrator in its package
inherits, and the residual misalignment of \S\ref{sec:horizon}.

\begin{table}[t]
\centering
\caption{The subjects: resolution sites per package under the criterion of \S\ref{sec:census},
machine-verified. Each site is recorded with its file, line and an anchor string,
and the census fails loudly if an anchor has moved, which it did twice during
construction. Downloads are the last-month figure from the same dated snapshot as
\S\ref{sec:census}'s denominator, queried \iwDate{}, and bound attention
rather than adoption. \orsec{F} re-queried the same denominator on \fsDate{} for the
index sweep and reports its own total for the audited ten from that later
snapshot, not this one}
\label{tab:census}
\small
\begin{tabular}{lrr}
\toprule
package (version pinned) & downloads/month & resolution sites \\
\midrule
\texttt{mapie 1.4.1} & 215,775 & 7 \\
\texttt{crepes 0.9.1} & 22,524 & 6 \\
\texttt{puncc 0.9.3} & 3,803 & 6 \\
\texttt{sktime 1.1.0} & 1,115,040 & 5 \\
\texttt{statsforecast 2.1.1} & 1,888,943 & 4 \\
\texttt{nonconformist 2.1.0} & 3,039 & 3 \\
\texttt{torchcp 1.2.1} & 460 & 3 \\
\texttt{darts 0.46.1} & 298,278 & 2 \\
\texttt{neuralforecast 3.2.1} & 299,178 & 2 \\
\texttt{river 0.25.0} & 214,218 & 1 \\
\midrule
\textbf{total} & \textbf{4,061,258} & \textbf{39} \\
\bottomrule
\end{tabular}

\end{table}

\paragraph{Outside the set it was built on.}
The packages below were chosen to be adversarial, not taken from what was to hand. All
\oosSearched{} on the unaudited side of \S\ref{sec:census}'s download denominator
were read against the criterion that section states, and every verdict is anchored to
a file and a line in a released sdist. \oosSite{} carry a site. \oosRefuted{} claimed
one on a first read and lost it on a second read made to break the claim:
\code{quantile-forest}'s expression is real and sits in \code{examples/}, which the
public-entry-point rule excludes, and \code{venn-abers} takes a rank over held-out
response labels rather than over scores. The remaining \oosNone{} resolve a level
against model output (quantile heads, a fitted inverse CDF, posterior draws) or,
in one case, against a remote service that returns the interval already built.
The suite of \S\ref{sec:suite} classifies \oosPackages{} of the \oosSite{} at pinned
versions, each in its own environment, and a fourth is measured by a separate
probe because its site is not a threshold function. \code{fortuna}'s split-conformal
regressors compute the required rank and then divide it by $n$ to recover a level, so
the level exceeds 1 wherever no rank can carry the request. At the interior size the
suite reads branch (\oosFortunaBranch{}), which \oosFortunaBranchName{}, landing on
\oosFortunaRank{} of \oosNok{} against a required \oosRequired{}; at \oosNbad{} the
method hands back
\oosFortunaBad{}, the largest calibration score, finite and with nothing said. Which
of those happens above a level of 1 belongs to the installed \code{jax}, not to
the library: the version pinned in that environment clamps where \code{numpy} raises.
\code{skforecast}'s public interval calibrator passes the requested level to
\code{numpy}'s default with no $(n+1)$ correction, which is branch
(\oosSkforecastBranch{}) at \oosSkforecastRank{} of \oosNok{}, \oosSkforecastBad{}
at \oosNbad{}, and no size at or below 2000 that delivers the level asked for. In
\code{crepes-weighted} two branches of one public method disagree:
the classic branch is \code{crepes}~0.9.1's expression, already in the census, and
lands on \oosCwClassicRank{} with a first delivering size of \oosCwClassicNmin{},
while the branch the fork adds lands on \oosCwWeightedRank{} at unit likelihood
ratios and delivers nothing until \oosCwWeightedNmin{}. Both err upward, so
that pair is a disagreement, not a shortfall.

\paragraph{A site whose consequence a clip decides.}
\code{conformal-tights} resolves a level over a held-out calibration split and
returns a per-quantile bias rather than a threshold, so the suite's shape does not fit
it and a separate probe runs the library and reads its own fitted state back. Across
two generators, four dataset sizes, both of its target types and the rails each
configuration resolves, \oosCtCells{} cells clear the library's own level-2 example
floor and fire. \oosCtClaim{} carry a one-sided coverage claim, and all \oosCtUnder{} of those
land short of the rank their own direction requires: at the smallest size clearing
that floor the resolved position is \oosCtHeadVirtual{} where level
\oosCtHeadLevel{} requires \oosCtHeadRequired{} of \oosCtHeadN{}. A coherence clip
then overrides the resolved value in \oosCtClipped{} of the \oosCtCells{},
\oosCtUnderClipped{} of them among the \oosCtUnder{} that undercover, leaving
\oosCtDecisive{} in which what the library keeps is what the level resolved to.
Delivered coverage alone would show no defect across much of that table while the
resolution underneath is wrong, which is why the rank landed on and the value kept
are separate columns. Below \oosCtFloor{} level-2 examples the block does not run at
all, and on a dataset of a thousand rows the library's default calibration split
leaves only \oosCtSkip{} of them.

This arm reports no negatives. A package searched and found to carry no
such site is not counted here, because a negative established by reading is not a
measurement, and no delivered interval coverage on real data is claimed for any of the
four.

\section{Delivered coverage on real data}
\label{sec:realdata}

Does the gap survive on real series, or only in exact arithmetic? Two forecasting
archives and two tabular suites, each shipped path measured against a paired arm off the
same fitted object. The forecasting libraries are \code{sktime}, \code{statsforecast}
and \code{darts}.

For each unit (a series, or a tabular dataset) we run the two arms of
\S\ref{sec:method}: the library as it ships, and the same library's scores, centre
and geometry at the required order statistic. We record coverage, interval width, and
the index figure arm A lands on. The discordance counts are the probe's own:
the \skEmpNinetyTwentyD{} difference at \code{sktime}'s setting-$20$ cell is
\skEmpNinetyTwentyDisc{} series of \skSeries{}, \skEmpNinetyTwentyLoss{} of them
reversed; the \daNinetyOneFiveD{} difference at \code{darts}' setting-$15$ cell is
\daNinetyOneFiveDisc{} of \skSeries{}. \code{sktime}'s setting-$20$ cell spans a median
\skEmpNinetyTwentyRank{} of \skEmpNinetyTwentyN{} where a median
\skEmpNinetyTwentyReq{} is required, so its deficit is several gaps and not one rank: each rail loses part of one
and the uncorrected $(n-1)$ scaling loses the rest.

\subsection{Forecasting}

Every forecasting cell here runs on the M1 monthly collection: the first \skSeries{}
series in archive order long enough for the arm's calibration settings, one held-out
point per series as the single test point, and a one-step horizon. That design is the
headline one because every arm runs it identically, so a cell's paired $\Delta$ is
comparable across libraries, though its absolute coverage is not (the base model
and the series still differ in what they leave to the residuals);
\S\ref{sec:robustness} rebuilds the two arms whose paired deltas carry the claim on
every eligible M3 series at \robOrigins{} rolling origins (\orsec{H}). The base forecaster
is a last-value naive model in the \code{sktime} and \code{statsforecast} arms and a
lag-1 linear regression in the \code{darts} arm, and the \code{statsforecast} rows are
its \code{conformal\_distribution} method; the other method it ships,
\code{conformal\_error}, delivers \sfErrDefaultNinety{} at the same default.

Table~\ref{tab:forecasting} reports the $0.90$ cells on the \skSeries{} monthly series.

\begin{table}[!htbp]
\centering
\caption{Delivered coverage at nominal $0.90$ on \skSeries{} real monthly series.
\emph{setting} is the library's own calibration
parameter, \code{n\_windows} for \code{statsforecast}, the initial window for
\code{sktime} and \code{cal\_length} for \code{darts}; $n$ is the number of
calibration scores that parameter produces: the setting itself for
\code{statsforecast} and \code{darts}, whose parameters are that count, and the
\emph{median} over the \skSeries{} series for \code{sktime}, whose parameter is a
window and whose series differ in length. Moving that parameter moves the base
forecaster's fit window, and so the scores themselves, not only the index arm A lands
on: that is why every claim in this paper is the paired difference and not arm A's
absolute column. Arm B is matched to arm A's geometry as well as to
its scores: \code{sktime}'s \code{empirical} resolves two levels on signed residuals,
so its arm B is the index pair spanning the required $\reqrank$ gaps and not one
threshold on absolute residuals, and the figure it lands on is a span. The last
column is a Wilson score interval on $\Delta$, computed from the cell's own
discordant count over its unit count and not from the standard error beside it
(\S\ref{sec:estimands}); on a coincidence cell that count is zero, so the interval
starts at zero and still has width. The \emph{pred.} column under arm A is the mean
over units of the position that unit's returned half-width recovers, over $n+1$, and
not the closed form of the shipped expression that the discussion below states; the
probe output also prints the coverage of the floor of that mean, which this table omits}
\label{tab:forecasting}
\scriptsize
\begin{tabular}{llrrrrrrrl}
\toprule
 & & & \multicolumn{2}{c}{arm A (shipped)} & & & & & \\
\cmidrule(lr){4-5}
library & setting & $n$ & measured & pred. & arm B & pred.\ $\Delta$ & $\Delta$ & s.e. & 95\% CI on $\Delta$ \\
\midrule
sktime \texttt{empirical} & 20 & 74 & 0.8920 & 0.8712 & 0.9280 & $+0.0369$ & $+0.0360$ & 0.0118 & [0.0191, 0.0670] \\
sktime \texttt{empirical} & 40 & 54 & 0.8880 & 0.8440 & 0.9520 & $+0.0404$ & $+0.0640^{\ddagger}$ & 0.0155 & [0.0398, 0.1014] \\
\midrule
statsforecast & 2 & 2 & 0.5840 & 0.6169 & vacuous & --- & $+0.4160^{\dagger}$ & 0.0312 & [0.3566, 0.4779] \\
statsforecast & 5 & 5 & 0.7200 & 0.7583 & vacuous & --- & $+0.2800^{\dagger}$ & 0.0285 & [0.2280, 0.3387] \\
statsforecast & 10 & 10 & 0.7800 & 0.8234 & 0.8880 & $+0.0857$ & $+0.1080$ & 0.0197 & [0.0753, 0.1526] \\
statsforecast & 20 & 20 & 0.8640 & 0.8613 & 0.9040 & $+0.0434$ & $+0.0400$ & 0.0124 & [0.0219, 0.0721] \\
statsforecast & 50 & 50 & 0.8760 & 0.8842 & 0.8880 & $+0.0178$ & $+0.0120$ & 0.0069 & [0.0041, 0.0347] \\
\midrule
darts & 10 & 10 & 0.8960 & 0.9091 & 0.8960 & $+0.0000$ & $+0.0000$ & 0.0000 & [0.0000, 0.0151] \\
darts & 15 & 15 & 0.8880 & 0.8750 & 0.9440 & $+0.0625$ & $+0.0560$ & 0.0146 & [0.0336, 0.0918] \\
darts & 30 & 30 & 0.8960 & 0.9034 & 0.8960 & $-0.0001^{\circ}$ & $+0.0000$ & 0.0000 & [0.0000, 0.0151] \\
darts & 35 & 35 & 0.8680 & 0.8889 & 0.9040 & $+0.0278$ & $+0.0360$ & 0.0118 & [0.0191, 0.0670] \\
darts & 50 & 50 & 0.8720 & 0.9020 & 0.8720 & $-0.0000^{\circ}$ & $+0.0000$ & 0.0000 & [0.0000, 0.0151] \\
darts & 55 & 55 & 0.8480 & 0.8929 & 0.8760 & $+0.0179$ & $+0.0280$ & 0.0105 & [0.0136, 0.0567] \\
\bottomrule
\end{tabular}

\end{table}

The $\Delta$ column does not hold one quantity throughout, and the
markers say which: $\dagger$ marks a cell where arm B is $+\infty$ for every unit, so
$\Delta$ is $1$ minus arm A's coverage and measures infeasibility rather than the
level-to-rank map; $\ddagger$ marks one where arm B is $+\infty$ for some units and
$\Delta$ is a mixture, for which the probe output carries the feasible-only difference.
Unmarked rows compare two finite bounds and are the ones the level-to-rank claim rests
on. The two predicted columns read one law. A threshold at fractional position $h$ of a
unit's $n$ scores delivers $h/(n+1)$ (pred.\ arm A) and falls
$(\reqrank-h)/(n+1)$ short of arm B (pred.\ $\Delta$). Both are means over the
same units, so on the \code{statsforecast} and \code{darts} blocks, where $n$ is one
number for every unit, the two columns sum to $\reqrank/(n+1)$ and a reader can check
one against the other. They do not sum on the two \code{sktime} rows, for two
reasons: $n$ there is a median over series of differing length, and the
$\ddagger$ row's arm B is $+\infty$ for part of its units. An interpolating helper returns a threshold strictly
between two scores and reaches the lower of them. A coincidence cell can print a
small negative while measured $\Delta$ is exactly zero; $\circ$ marks the cells where it
does, which are \S\ref{sec:severity}'s coincidence cells and not a sign error.

Three readings follow, the largest number last because it is not the level-to-rank map.

\textbf{(1) The design that isolates the cause.} \code{statsforecast} at $n=20$
delivers \sfDistNinetyTwoZeroA{} with a paired difference of $\sfDistNinetyTwoZeroD$
(s.e.\ \sfDistNinetyTwoZeroSE{}). Here the bound is feasible, both arms are finite, and
they differ only in the rank, by one, which the expression fixes before any series is read. \code{\_add\_conformal\_distribution\_intervals} stacks
the $n$ absolute scores with their negatives and hands the resulting $2n$ rows to
\code{np.quantile} at $1-\alpha/2$, under that function's default linear interpolation and
with no $(n+1)$ anywhere in the expression, so the upper cut lands at position
$\sfCutPosTwoZero{}$ of the $n$ scores where $\reqrank = \sfCutKstarTwoZero{}$ is
required. The floor of that position, the rank the cut reaches, is $0.9n$ at every $n$ divisible by ten and
$\reqrank$ is $0.9n+1$ there. Table~\ref{tab:forecasting}'s predicted arm A at this cell is
instead the mean recovered position, $\sfPosMeanTwoZero{}$, so dividing that column by
$n+1$ gives the realised mean, not $\sfCutPosTwoZero{}$. What $\Delta$ then counts is archive mass in one gap:
$\sfGapTwoZero{}$ of $\sfLedgerUnits{}$ test points fell between the two thresholds,
which is the $\sfDistNinetyTwoZeroD$ measured, against $\sfDistNinetyTwoZeroPred$
predicted. Sweeping every feasible size to $n=\sfSweepTo{}$ puts the three cells this
table runs among the mildest there are:
$\sfSweepShortOne{}$ of the $\sfSweepFeasible{}$ feasible sizes are one rank short,
every multiple of ten among them, and the remaining $\sfSweepShortTwo{}$ are short by
$\sfSweepMaxShort{}$.

\textbf{(2) Both methods land short, the default included.} \code{sktime} ships four
interval methods and this arm runs two. Its default, \code{empirical},
delivers \skEmpNinetyTwentyA{} and \skEmpNinetyFortyA{} at nominal $0.90$ for the
two calibration-window settings we run, with paired differences
$\skEmpNinetyTwentyD$ and $\skEmpNinetyFortyD$; its non-default \code{conformal} method is closer to
nominal at \skConfNinetyTwentyA{}. All \skCells{} cells the probe runs (both methods, both
calibration-window settings, both nominal levels) have a positive
paired difference; Table~\ref{tab:forecasting} prints the two under the default at
nominal $0.90$. The two methods also differ in geometry: the default resolves two levels on signed residuals, the other
one level on absolute ones, so each is paired against its own geometry and the two differences compare in
sign, not in size.

\textbf{(3) A shipped default that cannot be valid.} \code{statsforecast}'s
\code{ConformalIntervals} defaults to two calibration windows. At $n=2$ the
required rank for a $0.90$ interval is $\lceil 3 \times 0.90 \rceil = 3 > 2$, so
\emph{no} index into a two-element score set supports the requested level. The
library returns a finite interval regardless, and it delivers
\sfDefaultNinety{} against nominal $0.90$ on the first archive (and
\sfMThreeDefaultNinety{} on the second; \S\ref{sec:replication}); at nominal $0.95$ the same default
delivers \sfErrDefaultNinetyFive{}. This default is the largest shortfall we measure at either level, and
it is \textbf{infeasibility, not the level-to-rank
map}. Where the two can be separated, at $n = 10, 20, 50$, all above the floor, the paired difference
is $\sfDistNinetyOneZeroD$, $\sfDistNinetyTwoZeroD$ and
$\sfDistNinetyFiveZeroD$, shrinking with $n$ (Table~\ref{tab:forecasting}). The index the shipped
call lands on predicts every one of them: $\sfDistNinetyOneZeroPred$,
$\sfDistNinetyTwoZeroPred$ and $\sfDistNinetyFiveZeroPred$, within $\sfZWorst{}$
standard errors of the measurement at the worst of the three. Below the floor the same
index predicts $\sfDefaultNinetyPredA{}$ at $n=2$ against a measured
$\sfDefaultNinety{}$. On these series that is agreement, not attribution, and
\S\ref{sec:limitations} declines to attribute the raw-archive figure. The arm that
can attribute it is \S\ref{sec:attributable}'s, where the same default delivers
\atDefaultGot{} inside the bracket the landed index fixes.

\paragraph{A second archive.}
\label{sec:replication}
Every arm was rerun unchanged on a second, separate monthly collection, paired cell by
cell in \orsec{H}, where \code{statsforecast}'s default at $n=2$ delivers
\sfMThreeDefaultNinety{}. Each measured difference is positive except on the
coincidence cells of \S\ref{sec:severity}, which read exactly zero on both archives.
The arms are nested, so agreement in sign is weak evidence; what replicates with
something at stake is that every deficit cell reads positive rather than zero.

\paragraph{Absolute coverage, and which series.}
\label{sec:attributable}
Real series are not exchangeable, so an absolute miss on one is not attributable to the
convention (\S\ref{sec:limitations}). \orsec{H} keeps each donor series' scale and
residual shape and drops its ordering, by resampling its standardised residuals i.i.d.,
so Proposition~\ref{prop:interp} applies to an absolute number: measured coverage falls
strictly inside that proposition's bracket in \atStrict{} of \atCells{} cells, on
\atSeries{} independent series.

\label{sec:robustness}%
On every eligible series of M3 monthly, with \robOrigins{} rolling origins per series and
standard errors clustered by series, the sign of the paired difference agrees with the
\skSeries{}-series arm in \robSignAgree{} of \robCompared{} cells, and
\code{statsforecast}'s default still delivers \robSfDefaultNinety{}. The comparison
\S\ref{sec:estimands} says can fail does fail for \code{sktime}: in all \robSkCells{}
of its cells the measured $\Delta$ falls more than two clustered standard errors short
of pred.\ $\Delta$, while every feasible \code{statsforecast} cell sits within two. A
run on M1 monthly and a rerun on aligned residuals rule out how pred.\ $\Delta$ is
formed and the misaligned diagonal as the cause (\orsec{H}), which leaves
non-exchangeable one-step residuals as the explanation.

\subsection{Tabular: a null result, and why it is the important one}
\label{sec:tabular}

We repeated the design on tabular data: a classification suite and a
regression suite, with the selection rule fixed before any measurement and every
skipped dataset reported with its reason, across \tabConfigs{} implementations
resolving the same bound from the same scores on the same split
(Table~\ref{tab:tabular}).

\begin{table}[t]
\centering
\caption{Tabular arm. All \tabConfigs{} implementations reach the
required rank in every cell; \tabExactConfigs{} land exactly on it, the others one rank
above at the sizes where $\alpha(n{+}1)$ is a whole number, and the two-rail
configuration one gap wide at other sizes as well. The calibration sizes are
$n_{\text{cal}} \in \{\tabSizes\}$, at each level where the configuration is feasible.
At nominal $0.95$ no size in that grid makes $\alpha(n{+}1)$ a whole number, so a
single-rail row's one-rank-high case cannot show there and its exact verdicts at $0.95$
are untested at such sizes. Every departure
\emph{over}-covers. The \code{nonconformist} row's one departure is the harness's
$1-\text{coverage}$; handed the typed $\alpha$ the helper lands on the requirement in
\tabLitOn{} of \tabLitCells{} cells. \emph{(reference)} on the \code{torchcp} row names a helper that
indexes a whole-number rank with \code{kthvalue}; the level
$\lceil (N{+}1)(1-\alpha)\rceil/N$ it forms only guards the floor
(\code{torchcp/utils/common.py:68--69}) and reaches no quantile call. It is not an
oracle the other rows are scored against, and it is a different mechanism from
\emph{(CONTROL, exact)} above it, which appends $+\infty$ and inverts}
\label{tab:tabular}
\small
\begin{tabular}{lrl}
\toprule
implementation & cells & paired $\Delta$ \\
\midrule
\texttt{mapie SplitConformalClassifier (LAC)} & 10 & nonzero in 1/10; worst $-0.0195$ (3.3\,s.e.) \\
\texttt{mapie SplitConformalRegressor sym=True} & 10 & $0.0000$ in all 10 cells \\
\texttt{mapie SplitConformalRegressor sym=False} & 8 & nonzero in 3/8; worst $-0.0284$ (4.7\,s.e.) \\
\texttt{crepes ConformalRegressor.predict\_int} & 10 & nonzero in 1/10; worst $-0.0153$ (3.2\,s.e.) \\
\texttt{puncc BaseCalibrator (CONTROL, exact)} & 10 & $0.0000$ in all 10 cells \\
\texttt{torchcp calculate\_conformal\_value (reference)} & 10 & $0.0000$ in all 10 cells \\
\texttt{nonconformist AbsErrorErrFunc.apply\_inverse} & 10 & nonzero in 1/10; worst $-0.0153$ (3.2\,s.e.) \\
\bottomrule
\end{tabular}

\end{table}

All \tabConfigs{} implementations reach the required rank in
every cell they run; \tabExactConfigs{} of them return a paired difference of exactly
$\exactZero$ throughout, landing on it rather than above it. The gap between those two
counts is a width cost and not a validity one, and for the single-rail rows it is visible
only at the sizes where $\alpha(n{+}1)$ is a whole number. The calibration sizes this arm
ran before carried none of them, and over that grid those rows read exact. The asymmetric two-rail
configuration departs from exact most often.
It carries fewer cells than its neighbours because the library's own guard leaves its
smallest calibration sizes unusable at the tighter of the two nominal levels, $0.90$ and $0.95$.
It is exact in \tabExcExact{} of its \tabExcCells{} cells
and negative in the other \tabExcNonzero{}, worst case $\tabWorstDelta$, \tabWorstZ{}
standard errors. Negative, here, means arm A's rails span \emph{one gap more} than the
requirement: each rail resolves a corrected level and rounds outward, so the pair
over-covers, and arm B at the minimal sufficient span is the narrower interval. That is conservatism, and it must not be reported as undercoverage.

Which cells are exact is not arbitrary. At nominal $0.90$ it is exact at
$n_{\text{cal}} \in \{\tabExcExactSizes\}$ and one gap wide at
$n_{\text{cal}} \in \{\tabExcWideSizes\}$, which is
\S\ref{sec:severity}'s residue structure on the conservative side of the
requirement instead of the deficient one. The arithmetic that makes \code{darts}
exact on one residue class and one rank short off it makes this configuration exact on
one class and one gap wide off it.

\textbf{This null is a result.} What the two tables say jointly is narrower than
``conformal libraries undercover'' and better supported: the deficit concentrates in
implementations whose level-to-rank map is lossy, and three of the
tabular implementations reach the required rank by design, two of them by
never passing a level to a quantile function: one takes an integer rank via \code{kthvalue} with
an explicit $+\infty$ guard above the floor, one indexes the sorted scores
directly, and the third appends $+\infty$ to the scores, so an
\code{inverted\_cdf} quantile at the raw level lands on the requirement by
identity. Two of the three are among Table~\ref{tab:tabular}'s exact rows. The second, \code{nonconformist},
lands one rank long at the size where $\alpha(n{+}1)$ is a whole number, and that
departure is the harness's: the probe passes it $1-\text{coverage}$, which in doubles sits
just below $\alpha$, and the helper floors that value times $n+1$, which there falls one
short of the whole number. Handed the $\alpha$ a caller types, the same helper lands on
the requirement in \tabLitOn{} of the grid's \tabLitCells{} cells. The exact rows are
\code{torchcp}, \code{puncc} and \code{mapie}'s symmetric regressor.

\paragraph{A control that must read zero.} One of the \tabConfigs{} is exact by
construction, so a defect reported there would be the harness measuring itself. It
reads $\exactZero$ in every cell.

\paragraph{A streaming estimator.}
\code{river}'s \code{RegressionJackknife} (version \rvVersion{}) resolves its bound
through an approximate streaming quantile estimator with no finite-sample correction
and no warm-up guard, branch~(g). Driven with a regressor that predicts zero, so that
its residuals are the i.i.d.\ standard-normal targets themselves, at a requested
\rvRequested{} and over \rvReps{} replications per size, it delivers \rvCovThree{} at
three calibration points, \rvCovTen{} at ten, \rvCovFifty{} at fifty and
\rvCovTwoHundred{} at two hundred, and \rvCovThousand{} at a thousand. Below a few dozen
points the estimator's own error, not the level-to-rank map, sets the gap, and coverage
is not monotone in $n$; no paired arm can isolate the rank here, because the helper
never forms an order statistic to compare against.

\subsection{Severity is not branch membership}
\label{sec:severity}

Knowing which branch a library takes does not tell a caller how much coverage they lost.
This section measures the miss where nothing but the rank varies.

\code{darts} and \code{sktime} both omit the correction and behave very
differently, entirely because of the level-to-rank map: \code{darts}
passes an uncorrected level through \code{method="higher"}, which returns an actual
order statistic; \code{sktime} passes two uncorrected levels through linear
interpolation on \emph{signed} residuals, which returns a value between order
statistics on each rail.

\paragraph{The residue structure.}
For \code{darts}, the delivered rank is
$\delrank = \lceil c(n-1) \rceil + 1$ with $c$ the requested coverage, passed uncorrected as a two-sided range,
against a required $\reqrank = \lceil c(n+1) \rceil$. The deficit
$\reqrank - \delrank$ is not erratic: over $n = 9,\dots,60$ it is $0$ at exactly
\dmZero{} values of $n$ and $1$ at the other \dmOne{}, and it is never $2$ at any
$n$, since $\reqrank - \delrank = \lceil c(n+1) \rceil - \lceil c(n-1) \rceil - 1 < 2c < 2$
for $c < 1$
(\orsec{H} plots it). The zeros are arithmetic
coincidences, not correctness.

\begin{figure}[t]
\centering
\includegraphics[width=0.85\textwidth]{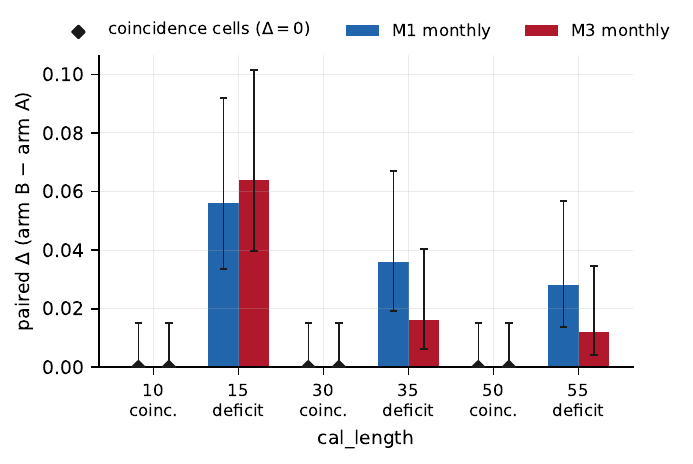}
\caption{Paired difference per calibration length on two archives. Coincidence
cells are exactly zero by construction and act as the harness's negative control.
Error bars are Wilson 95\% intervals on the paired difference
(\S\ref{sec:estimands})}
\label{fig:darts}
\end{figure}

A study that samples
calibration sizes landing on those coincidence cells will report zeros and conclude
the library is correct; one that samples the complement will report a uniform
one-rank deficit. We therefore sample both bands deliberately, three cells each,
and the paired differences separate cleanly: exactly $\exactZero$ on the
\daCoincCells{} coincidence cells and between $\daDeficMin$ and $\daDeficMax$ on
the \daDeficCells{} deficit cells. The probe runs the same lengths at nominal $0.95$ too,
where the rule puts every feasible length in the deficit band and none on a
coincidence; those cells are printed in the probe output only, and no claim here rests
on them. The second archive replicates the split, zero on
every coincidence cell and positive on every deficit cell, but not the range: there the
deficit cells read $\daMThreeDeficMin$ to $\daMThreeDeficMax$ (Fig.~\ref{fig:darts}). The coincidence cells are the negative control: there both
arms select the same order statistic, so their zero is fixed by construction and
tests the harness, not the convention. The evidence is the deficit cells, on the same
library, data and code path, where $\Delta$ is positive and sits near the pred.\
$\Delta$ the landed index gives (Table~\ref{tab:forecasting}).

What the repair costs in width on these six cells, and an instrumented decomposition
that separates the map from in-sample residual bias, are in \orsec{H}.

\section{Three defects in one multi-horizon helper}
\label{sec:horizon}

A helper whose delivered coverage sits near nominal can still be wrong in three ways.
This section separates the three defects in one such helper. Two of them point in
opposite directions and partly cancel.

The largest shortfall in \S\ref{sec:realdata} is infeasibility at a two-window
default, and the natural objection is that this is a corner: nobody calibrates on
two points, and $n \ge \rmTheoreticalNinety{}$ is cheap. This section shows why the
objection fails. It also reports a defect we were not looking for, found when a
check written to establish one thing rejected it.

\subsection{A residual-alignment off-by-one}
\label{sec:align}

To measure absolute coverage we need exchangeable scores, so we generated random
walks: a last-value forecaster's one-step residuals are then the innovations
themselves. We wrote a \emph{structural} check (the score set the library resolves its level from must be
the absolute first differences of the input), and it failed.

Solving each entry of \code{sktime}'s \code{residuals\_matrix\_} for the index pair
that produces it shows why. The matrix holds
$A[i,j] = y_j - y_{o_i - 1}$ for origin $o_i$, so its offset-$0$ diagonal is
\emph{already} the one-step residual set. The library slices it with
\code{resids = np.diagonal(residuals\_matrix, offset=offset)} where \code{offset} is
the relative horizon, so a step-$\ell$ forecast is calibrated on the offset-$\ell$
diagonal, which holds $(\ell{+}1)$-step residuals ($\ell$ is the forecast step,
not the virtual index $h$; the filed report in \orsec{I} calls the step $h$).

We establish this twice, because it is a claim about someone else's code and this
programme has retracted two of those. Structurally, the index pairs on the
offset-$k$ diagonal span exactly $k+1$ steps, asserted for $k = 0,1,2$.
Statistically, on a random walk with unit innovation variance the variance of the
offset-$k$ diagonal must be $(k{+}1)\sigma^2$ if the off-by-one is real and
$k\sigma^2$ if it is not; over \hzAlignReps{} series the offset-$0$, $1$ and $2$
diagonals measure \hzVarZero{}, \hzVarOne{} and \hzVarTwo{} (standard errors
\hzVarZeroSe{}, \hzVarOneSe{} and \hzVarTwoSe{}). Because each diagonal's own mean is
subtracted, the $(k{+}1)$ hypothesis predicts slightly less than $k+1$:
\hzVarZeroExp{}, \hzVarOneExp{} and \hzVarTwoExp{}. Each measurement is within two
standard errors of that expectation; \orsec{I} sets them against both predictions.

The direction is why this has survived. For an integrated series a
longer-horizon residual is larger, so the interval is too \emph{wide}: the defect
is conservative, and a conservative interval fails no coverage test. It is
orthogonal to everything else in this paper: an index error in the score set, not a
level-to-rank error, and it points the opposite way, so in the shipped helper the two
partly cancel.

One consequence for our own method: because the shipped scores are
$(\ell{+}1)$-step, they are \emph{not} exchangeable with the test residual at any
horizon, whatever the data. Absolute coverage through this public API is therefore
not attributable to any single cause, and we run four arms instead of one.

\subsection{The floor scales with whatever divides $\alpha$}

The floor $n \ge 1/\alpha - 1$ is a property of the level actually
\emph{resolved}, not of the level requested. The span identity of \orsec{K} is the
general statement: a construction resolving $1-\alpha_1,\dots,1-\alpha_L$ with
$\sum_j\alpha_j = \alpha$ is feasible exactly for $n \ge 1/\min_j\alpha_j - 1$, so
$n+1$ is multiplied by $\alpha/\min_j\alpha_j$ and not by the number of pieces. A Bonferroni split
across a horizon of length $H$ has $\alpha_j = \alpha/H$ throughout, so the floor
becomes $n \ge H/\alpha - 1$ (Table~\ref{tab:horizonfloor}); the same proposition at
$L = 2$ gives the two-rail floor $n \ge 2/\alpha - 1$ that \S\ref{sec:realdata}'s arm B
is built to respect, and at an \emph{uneven} split it gives something worse than
either. At $H=12$ and nominal $0.95$
that is \hzFloorTwelveNinetyFive{} calibration windows, a factor \hzRatioTwelve{}
above the single-step floor; at $H=24$ it is \hzFloorTwentyFourNinetyFive{}; a
weekly horizon of a year, $H=52$, needs \hzFloorFiftyTwoNinetyFive{}, a factor
\hzRatioFiftyTwo{}. This claim needs no exchangeability: whether an index into a score set
of size $n$ can support level $1 - \alpha/H$ is arithmetic.

\begin{table}[!htbp]
\centering
\caption{The Bonferroni floor. Exact arithmetic; ``vs $H{=}1$'' is the multiple of
the single-step floor of \S\ref{sec:background}}
\label{tab:horizonfloor}
\small
\begin{tabular}{rrrrr}
\toprule
nominal & $H$ & $\alpha/H$ & floor on $n$ & vs $H{=}1$ \\
\midrule
0.90 & 1 & 0.10000 & 9 & 1.0$\times$ \\
0.90 & 3 & 0.03333 & 29 & 3.2$\times$ \\
0.90 & 6 & 0.01667 & 59 & 6.6$\times$ \\
0.90 & 12 & 0.00833 & 119 & 13.2$\times$ \\
0.90 & 24 & 0.00417 & 239 & 26.6$\times$ \\
0.90 & 52 & 0.00192 & 519 & 57.7$\times$ \\
\midrule
0.95 & 1 & 0.05000 & 19 & 1.0$\times$ \\
0.95 & 3 & 0.01667 & 59 & 3.1$\times$ \\
0.95 & 6 & 0.00833 & 119 & 6.3$\times$ \\
0.95 & 12 & 0.00417 & 239 & 12.6$\times$ \\
0.95 & 24 & 0.00208 & 479 & 25.2$\times$ \\
0.95 & 52 & 0.00096 & 1039 & 54.7$\times$ \\
\bottomrule
\end{tabular}

\end{table}

\paragraph{Three separated failures.}
\code{sktime}'s \code{conformal\_bonferroni} passes an uncorrected level, divided by the
horizon length, to \code{numpy.quantile} at \code{linear}, on a misaligned score set.
Four arms off one fit separate the three defects (\orsec{D}). \hzInfeasCells{} of
\hzCells{} cells sit below $H/\alpha - 1$ and still return finite intervals. At $H{=}1$
misalignment makes the shipped interval \emph{over}-cover, \hzOneA{} against
\hzOneD{} with the scores aligned, which hides a level-to-rank cost of up to
\hzRankCost{}. The checklist of \S\ref{sec:suite} asks for the level resolved and the
size it was resolved from for this reason.

\section{The feasibility floor has no shared contract}
\label{sec:boundary}

What does a library hand back when no rank carries the requested level? Every path this
audit can reach, executed one calibration size below the floor.

Everything measured so far is worth a bounded number of ranks: one or \sfSweepMaxShort{} for a helper
resolving a single level (\S\ref{sec:realdata}), and one per rail plus the scaling for a helper resolving two
(\S\ref{sec:realdata}). All of it is
$O(1/n)$. This section is about the calibration sizes where the guarantee does not
exist at all.

Below the floor $n < 1/\alpha - 1$ there is no rank in the sample whose coverage
reaches the requested level. A helper that returns a number there is not slightly
wrong. It is reporting a guarantee that no index into its own score set can carry.
Two results above say the region is ordinary: the largest
shortfall in \S\ref{sec:realdata} is a \emph{shipped default} sitting inside it, and
\S\ref{sec:horizon} raises the floor to $H/\alpha - 1$, so a Bonferroni band
at $H{=}12$ and nominal $0.95$ needs \hzFloorTwelveNinetyFive{} calibration windows
and no shipped default approaches that.

The conformance suite probes every helper at a
size one below the floor for a requested $0.90$, and
the \emph{at $n$ below feas.} column of Table~\ref{tab:conformance} records what each does.

The denominator is not the census's. A row of
Table~\ref{tab:conformance} is a \emph{call path}: \confPaths{} paths over
\confSites{} distinct resolution sites, the difference being the \confSharedPaths{}
site \S\ref{sec:unbounded} reaches twice, once on each side of a public keyword. Each
row carries two verdicts, the helper's in the \emph{at $n$ below feas.} column and the
shallowest public route's in the last, and the abstract's public call paths are the
last column's. The counts below are over paths, the population that can be executed at the
floor. Neither number is a proportion of the ecosystem and we do not present one.

\paragraph{How \cenPTwo{} sites become \confPaths{} rows.} Sites and table rows
count different populations, and setting them side by side needs the join. The census
splits by disposition, $\cenPTwo{} = \cenDriven{} + \cenShared{} + \cenPvalue{} +
\cenAbsent{}$: driven sites the suite constructs and calls, sites sharing one bare
\code{numpy.quantile} call at an uncorrected level that the table carries once per
definition rather than once per caller, $p$-value denominators with no threshold to
land on, and sites outside this suite's reach. That last figure is a statement about
the suite and not about those sites: we give no reason for each, because a reason
nobody measured is not a finding. The \cenAbsent{} outside the suite's reach and the
\cenPvalue{} $p$-value sites, \cenAbsent{} + \cenPvalue{} of \cenPTwo{} in all, carry a
branch letter read from source rather than run, RQ1's own caveat about that method
(\orsec{J}). The first two terms are the \cenExecuted{} classified by execution: the
\cenDriven{} driven sites through their own expressions, the \cenShared{} shared ones
through the \code{numpy} method they share and not through their own level
expressions. \orsec{J} tabulates that split and the arithmetic that closes it to
\confPaths{}, carries the same join site by site, and puts it beside the article's four
other pairs of counts.

\begin{itemize}\itemsep1pt
\item \textbf{\confBelowFinite{} of \confPaths{} return a finite threshold}, and a
      further \confBelowCount{} returns a raw count that a caller could mistake for
      one. \textbf{None of those \confBelowSilent{} emits a warning.}
\item \confBelowInf{} return $+\infty$, and all \confBelowInf{} warn.
\item \confBelowRaise{} raise.
\end{itemize}

Two things follow that a per-library reading does not give. The first is the
\confBelowSilent{}: every path we can execute below the floor that hands back a
number without warning. All but one of the \confBelowFinite{} deliver the sample
maximum: rank $n$, so $n/(n+1)$, \emph{whatever level was asked for}. The
exception is the shared \code{numpy} \code{linear} path
(Table~\ref{tab:conformance}): an interpolated value strictly below the sample
maximum, so a strictly worse guarantee than the max would give. The further
\confBelowCount{} path returns a raw count rather than a threshold (at $n=8$ it returns $\lacCountAtBad$, above
every score), the opposite direction of harm. At
$n=2$ and a requested $0.95$ the worst sample-maximum delivery is
$\clipWorstDelivered{}$.

The second is that the \confBelowRaise{} and the \confBelowInf{} disagree with each
other. Raising and returning $+\infty$ are both defensible and they are not the same
contract: one makes the infeasible configuration a programming error, the other makes
it a representable answer a caller can propagate. A practitioner switching libraries
at a fixed $n$ and level can therefore get an exception, an infinity, or a finite
number, from code implementing the same textbook construction. There is no shared
contract at the floor to violate. The finding is not that particular helpers are
wrong, but that the floor is the one place where the ecosystem never agreed what
correct means.

A count over helper invocations is not a count over routes a caller has, and here the
two differ. Every row but the two \code{numpy} rows was called again from the
shallowest entry point its library documents, at the same $n=8$, the two \code{numpy}
rows through one consumer each (\code{statsforecast}'s \code{ConformalIntervals} and
\code{darts}' \code{ConformalNaiveModel}), and both return a finite interval in
silence there, as their real-data runs below the floor already did
(\S\ref{sec:realdata}, \S\ref{sec:severity}). The verdict is the table's last column:
\confRouteSilent{} of \confPaths{} hand a caller a finite number in silence,
\confRouteRaise{} refuse before the helper runs, and the remaining \confRouteInf{} surface
$+\infty$ with a warning (\code{torchcp}'s at the threshold, which its interval
constructor then fails on; the route tried is \code{calibrate}, so that later failure
is not tallied as a raise). What separates the two counts is where the guard sits.
\code{puncc} raises \emph{Alpha is too small} at each of its three high-level
predictors, and \code{mapie}'s regressor and classifier entry points raise on the
sample count, so the clamped expressions those rows name are unreachable from a
documented call, except through the keyword of \S\ref{sec:unbounded}, which removes
the guard and then returns the finite bound in silence. Above
\code{nonconformist}'s interval predictor, the two \code{numpy} consumers and
\code{statsforecast}'s seasonal pool there is no such guard. The two numbers answer
different questions: \confBelowSilent{} is how many
shipped expressions behave this way, \confRouteSilent{} how many a caller can reach,
and a maintainer reading a defect report asks the second one first.

\subsection{A finite bound where the caller asked for an infinite one}
\label{sec:unbounded}

One documented keyword promises an infinite bound and removes the guard that would
deliver it. \code{mapie}~1.4.1 exposes \code{allow\_infinite\_bounds} on
\code{SplitConformalRegressor.predict\_interval}. A caller sets it to be given
$\pm\infty$ where no finite \emph{deterministic} bound is valid, branch (c), the option \orsec{G} keeps as the honest
deterministic one. Below the feasibility floor it returns a finite interval built on
the sample maximum instead.

The mechanism is three lines of the pinned source. At \code{mapie/regression/regression.py:1714} the
flag decides whether the calibration-size guard runs at all, so setting it removes
the guard rather than changing what happens past it. \code{interface.py:137} forms
the corrected level $\lceil \alpha_{\text{ref}}(n{+}1)\rceil/n$, where \code{mapie}'s
$\alpha_{\text{ref}}$ is the coverage level, $1-\alpha$ in this paper's notation, and
\code{interface.py:138} clips it into $[0,1]$. The branch that would return infinity,
at \code{interface.py:150}, is written against $\alpha_{\text{ref}}$, the
uncorrected level, not the corrected one, while infeasibility is the condition
$\lceil \alpha_{\text{ref}}(n{+}1)\rceil > n$: satisfied at values of
$\alpha_{\text{ref}}$ well short of $1$. So in exactly the regime the flag exists
for, the infinity branch does not fire, the clip brings a corrected level above $1$
back down to $1$, and the quantile call at level $1$ under \code{method="lower"}
returns the sample maximum. The guard and the branch it guards are written against
two different quantities.

\paragraph{What it delivers.}
On a tie-free score set the returned threshold is the rank it landed on, and at
every infeasible size that rank is $n$. Delivered coverage is therefore $n/(n+1)$,
and it does not depend on the level requested. We enumerate the window from $n = 2$,
the smallest size the probe takes, rather than sample it: \clipCells{} infeasible sizes over \clipLevels{} levels, exact
arithmetic, with the first feasible size at each level carried as a control where
the maximum \emph{is} the correct answer and the shortfall must read zero
(Table~\ref{tab:clip}). At
$n = \clipN{}$ and nominal \clipLevel{}, where the required rank is \clipReq{} of
\clipN{}, the flagged path delivers \clipDelivered{}: a shortfall of
\clipShortfall{}. The window closes at $n = \clipWindowTop{}$ and the shortfall
grows as the calibration set shrinks, reaching \clipWorstShortfall{} at
$n = \clipWorstN{}$.

\begin{table}[t]
\centering
\caption{The infeasible window from $n = 2$ the flagged path returns a finite bound on,
both levels, exact arithmetic. ``delivered'' is $n/(n{+}1)$ at the rank the flagged
path lands on; ``shortfall'' is the requested level minus delivered, as the probe prints it. The marked row at each
level is the first feasible size, carried as a control}
\label{tab:clip}
\begin{tabular}{rrrr}
\toprule
$n$ & required rank & delivered & shortfall \\
\midrule
\multicolumn{4}{l}{\textit{nominal 0.90}} \\
2 & 3 & 0.6667 & +0.2333 \\
3 & 4 & 0.7500 & +0.1500 \\
4 & 5 & 0.8000 & +0.1000 \\
5 & 6 & 0.8333 & +0.0667 \\
6 & 7 & 0.8571 & +0.0429 \\
7 & 8 & 0.8750 & +0.0250 \\
8 & 9 & 0.8889 & +0.0111 \\
9 & 9 & 0.9000 & +0.0000$^{\dagger}$ \\
\addlinespace
\multicolumn{4}{l}{\textit{nominal 0.95}} \\
2 & 3 & 0.6667 & +0.2833 \\
3 & 4 & 0.7500 & +0.2000 \\
4 & 5 & 0.8000 & +0.1500 \\
5 & 6 & 0.8333 & +0.1167 \\
6 & 7 & 0.8571 & +0.0929 \\
7 & 8 & 0.8750 & +0.0750 \\
8 & 9 & 0.8889 & +0.0611 \\
9 & 10 & 0.9000 & +0.0500 \\
10 & 11 & 0.9091 & +0.0409 \\
11 & 12 & 0.9167 & +0.0333 \\
12 & 13 & 0.9231 & +0.0269 \\
13 & 14 & 0.9286 & +0.0214 \\
14 & 15 & 0.9333 & +0.0167 \\
15 & 16 & 0.9375 & +0.0125 \\
16 & 17 & 0.9412 & +0.0088 \\
17 & 18 & 0.9444 & +0.0056 \\
18 & 19 & 0.9474 & +0.0026 \\
19 & 19 & 0.9500 & +0.0000$^{\dagger}$ \\
\addlinespace
\bottomrule
\end{tabular}

\smallskip
\noindent\footnotesize $^{\dagger}$the first feasible size at this level, carried as the
control: the flagged path returns the correct answer here, so its shortfall must read zero.

\end{table}

Measured end to end through \code{predict\_interval} on i.i.d.\ draws with a prefit
base model, over \clipReps{} replications per cell, the same $n = \clipN{}$ cell
covers \clipMeasured{} with a standard error of \clipSE{}; no cell sits further than
\clipMaxZ{} standard errors from its exact $n/(n+1)$. Every interval returned was
finite. On the default path the same call raises.

\paragraph{Why this is not the one-rank deficit.}
Everywhere else in this audit the defect is a threshold a bounded number of order
statistics below the required one (one or \sfSweepMaxShort{} for a single-level helper, a few gaps of span for a
two-rail one) and therefore $O(1/n)$. Here no order statistic in the sample
reaches the requested level, so the deficit is the whole distance between the top of
the sample and a guarantee the sample cannot carry. It is also
the only defect we measure that selects for the careful caller: the flag's documented
purpose is to permit an infinite bound, so the callers who set it are exactly the
ones who wanted to be told, and they are the ones not told.
\code{TimeSeriesRegressor.adapt\_conformal\_inference}, a public method, passes it as
\code{True} in its own update loop (\code{time\_series\_regression.py:319}), so a
caller who never sets the keyword reaches the path through an ordinary API call.

\paragraph{How wide the door is.} One flag on one method would be a narrow finding,
so we counted. Reading the pinned package's signatures with \code{ast} instead of by
eye, \clipSigs{} carry the keyword and \clipPublic{} of those sit on a public class
and a public method. Constructing every public regressor the probe can build and
driving each to the same infeasible configuration, \clipReaches{} of \clipTried{}
raise on the default path and return a finite interval on the flagged one. The other
three of the \clipPublic{} are not driven end to end here, and the probe names rather
than drops them: \code{BaseRegressionScore.get\_bounds} is the shared bound builder and
not a regressor class, \code{ConformalizedQuantileRegressor.predict\_interval} is the
quantile-regression path, and \code{TimeSeriesRegressor.predict} takes the keyword as its own
argument, default \code{False} (\code{time\_series\_regression.py:417}), and is
reached with it set \code{True} from \code{adapt\_conformal\_inference} (\code{:319}). The
behaviour lives in the shared bound-construction layer, which the classes above it
inherit rather than implement.

The site itself is not new: it is one of the \cenPTwo{} that \S\ref{sec:census}
anchors, classified there as branch~(b), the clamp. The flag adds no site, since one
expression still turns one level into one index, the criterion of
\S\ref{sec:census}, and it does not change which branch that expression reaches:
Table~\ref{tab:conformance} carries the same site twice, once on each path, in
branch~(b) both times. A caller asking for branch~(c), $\pm\infty$, gets branch~(b)
and no warning.

This finding was filed upstream, like every other defect this paper reports\deanon{,
as \code{mapie} Issue~\#980; it was closed when a fix written by another contributor
(\#987) merged, and that fix shipped in a release later than the one pinned here, so
what this section measures is the behaviour at the pinned version}{}. The scan that
produced it is \deanon{in the artifact}{in the artifact, which this review copy
withholds}.

\section{What this leaves behind, and how it was checked}
\label{sec:suite}

A maintainer who accepts the diagnosis needs a test that catches it and a way to report
what their call resolves. This section supplies both, and which calibration sizes the
test must probe is decidable rather than a matter of taste.

The census is a snapshot; the interfaces are not, so we package the diagnosis. Given
any callable mapping scores and a level to a threshold, the
suite reports the branch, the rank landed on, the delivered coverage as
$\delrank/(n+1)$, the smallest $n$ at which the requested level is delivered, and
whether the helper warns at the floor (Table~\ref{tab:conformance}). The \code{puncc}
row's $n_{\min}$ is the harness's: it passes $1-\text{coverage}$, which in floating point
falls just below the typed $\alpha$, and \code{puncc}'s guard refuses it at the floor
itself, where the typed $\alpha$ passes; Table~\ref{tab:tabular}'s
\code{nonconformist} row departs for the same reason. The suite
validates itself at import by classifying reference implementations of every
branch, so it fails instead of mislabelling if it drifts.

\begin{table}[!htbp]
\centering
\caption{Conformance results for the shipped helpers, at requested level
$0.90$. \code{numpy}'s default delivers at no $n \le 2000$. On a path that passes its level uncorrected, the smallest
delivering $n$ heads a residue class rather than opening a range, so collecting more
calibration data is not the remedy (item~\ref{itm:residue} of \S\ref{sec:imported}).
\confWarns{} of
\confPaths{} call paths warn at the floor. The ``at $n$ below feas.'' column is what
each path returns at $n=8$, one size under the floor, on the tie-free score set
$1,\dots,n$: a value there \emph{is} the rank the helper selected, and a non-integer
one means it interpolated between two order statistics. \code{raises} and $+\infty$
are behaviours rather than values, and the \emph{not a threshold} row is the raw
count of \S\ref{sec:census}, which is why it stands above every score; its delivered and
$n_{\min}$ entries read that count as a rank and are not a coverage. The \code{linear} row
is the bare call at $0.90$: \code{statsforecast} and \code{neuralforecast} reach it on a
stack of $2n$ signed rows cut at $1-\alpha/2$, whose landed position \orsec{H} gives.
\code{crepes ConformalRegressor} delivers from $n=10$ and not $9$ because it indexes
\code{int((1-confidence)*(n+1))-1} (\code{crepes/base.py:900}), and $1-0.9$ evaluates just
below one tenth in floating point, so the error is conservative. The rows are paths, not sites:
\confSharedPaths{} site appears twice, on the two sides of a public keyword, so the
table holds \confSites{} distinct sites. A two-rail helper resolves
$1-(1-\text{coverage})/2$ on each rail, so its \emph{delivered} entry is that per-rail
level and its floor is the per-rail floor. That is why
\code{crepes ConformalPredictiveSystem} reads a higher delivered level and a larger
$n_{\min}$ than its neighbours under one requested $0.90$; the probe output says so,
and the row is not a disagreement. The
\emph{public route} column asks the same question of the shallowest entry point each
library documents, at the same $n=8$: \code{finite} means a caller reaches a finite
number from a documented call, \code{raises} that every entry point tried refuses
before the helper runs; the entry points tried are printed per row in the
probe output. The two \code{numpy} rows are called through
one consumer each, statsforecast's \code{ConformalIntervals} and darts'
\code{ConformalNaiveModel}, at the same $n=8$ and $0.90$}
\label{tab:conformance}
\scriptsize
\begin{tabular}{@{}>{\raggedright\arraybackslash}p{0.30\textwidth}lrrrcl@{}}
\toprule
helper & branch & at $n$ below feas. & delivered & $n_{\min}$ & warns & public route \\
\midrule
\texttt{numpy method=\textquotesingle{}linear\textquotesingle{} (sktime, statsforecast, neuralforecast)} & (d) no correction & 7.300 & --- & $>$2000 & --- & finite \\
\texttt{numpy method=\textquotesingle{}higher\textquotesingle{} (darts)} & (d) no correction & 8.000 & 0.9000 & 9 & --- & finite \\
\texttt{statsforecast \_oriented\_index} & (b) clamps & 8.000 & 0.9000 & 9 & --- & finite \\
\texttt{mapie get\_quantile} & (b) clamps & 8.000 & 0.9000 & 9 & --- & raises \\
\texttt{mapie get\_quantile [allow\_infinite\_bounds=True]} & (b) clamps & 8.000 & 0.9000 & 9 & --- & finite \\
\texttt{crepes ConformalRegressor.predict\_int} & (c) $+\infty$ & +inf & 0.9091 & 10 & yes & +inf, warns \\
\texttt{puncc BaseCalibrator.compute\_quantile} & (a) raises & raises & 0.9091 & 10 & --- & raises \\
\texttt{torchcp calculate\_conformal\_value} & (c) $+\infty$ & +inf & 0.9000 & 9 & yes & +inf, warns \\
\texttt{mapie LAC quantiles [prefit/mean -> delegates]} & (a) raises & raises & 0.9000 & 9 & --- & raises \\
\texttt{mapie LAC quantiles [cv=5/crossval -> count scale]} & not a threshold & 8.100 & 0.9000 & 9 & --- & raises \\
\texttt{mapie \_beta\_optimize + get\_quantile (composed)} & (b) clamps & 8.000 & 0.9000 & 9 & --- & raises \\
\texttt{crepes ConformalPredictiveSystem.predict\_int} & (c) $+\infty$ & +inf & 0.9500 & 19 & yes & +inf, warns \\
\texttt{puncc api/utils.py quantile (shared utility)} & (d) no correction & 8.000 & 0.9000 & 9 & --- & raises \\
\texttt{nonconformist AbsErrorErrFunc.apply\_inverse} & (b) clamps & 8.000 & 0.9000 & 9 & --- & finite \\
\bottomrule
\end{tabular}

\end{table}

\paragraph{Which calibration sizes the suite probes.}
\label{sec:certificate}
A fixture pinned to a coincidence cell cannot discriminate, and items~\ref{itm:residue}
and~\ref{itm:cert} of \S\ref{sec:imported} turn that into a certificate: \ceCertSize{}
sizes, computed from the level under test, separate all \ceClasses{} classes among the
\ceRules{} stated candidate rules. Outside the candidates the certified sizes are a screen
and not a proof; the complete check, its blind spot and both readings of the projects'
own fixtures are in \orsec{C} and \orsec{J}.

\paragraph{Three items to report.}
We propose that any paper or library claiming a distribution-free coverage level
report: \textbf{(1)} the delivered coverage, not the requested one, as the rank
landed on over $n+1$; \textbf{(2)} that rank, as $k$ of $n$, so a reader can check
the arithmetic without rerunning anything; and \textbf{(3)} the calibration set
size, together with whether \emph{any} valid finite bound exists at that size for
that level. Item (3) is the one that catches a finite interval returned where none
is possible, which is the failure mode of \S\ref{sec:realdata}'s largest cell.

\subsection{How this was checked, and what was filed}
\label{sec:checks}

Which of these findings reached the people who own the code, and what happened next? The
trackers answer that.

Library tests do not catch this for three reasons that generalise past conformal
prediction. An oracle sharing the helper's own convention is a consistency check,
not an independent one. A fixture at a coincidence cell cannot discriminate: by
\S\ref{sec:severity} an uncorrected rank equals the required rank on a residue
class, so a test pinned to such an $n$ passes forever, the coincidental correctness
of \citet{masri2014} with the faulty expression executed on every run. And the
failure is one rank, not an order of magnitude: a tolerance loose enough for Monte
Carlo noise hides an $O(1/n)$ deficit.

\paragraph{\rfcBelowFloor{} of \rfcFixtures{} calibration fixtures sit below the
floor.} That last reason is measurable in the projects themselves: of the \rfcFixtures{} scorable
calibration sizes these projects pin in their own tests, \textbf{\rfcBelowFloor{} sit
below the feasibility floor for $0.90$}, where no rank carries the level at all, so
what they pin is behaviour outside the region the guarantee is defined in. Every fixture
is scored against the floor for $0.90$; the level each test itself requests was not read,
so a fixture whose own test asks for a lower level may sit above that test's floor.
\orsec{J} gives the parsing rules, the per-project counts and how far the fixtures above
the floor sit from the certificate's discriminating power.

Every defect this paper reports as
a defect was filed upstream before this submission:
\deanon{issue and pull-request numbers per site in \orsec{I}'s disclosure table,
every filing predating this submission}{dates, packages, sites and defect
descriptions in the anonymised manifest accompanying this submission
(\code{DISCLOSURE-review.md}), report numbers withheld and supplied to the Action
Editor as a confidential comment, inlined at camera-ready}, spanning five
packages. Absence from that
record is not a claim of correctness: the count sits far below the \cenPTwo{}
sites of the census, filed only where what a library ships and what it says
disagree and not when the behaviour is deliberate and described, however lossy,
so \code{statsforecast}'s \sfDefaultNinety{} default
(Table~\ref{tab:forecasting}) is answered by \orsec{G}'s construction
rather than a report. Queried against the trackers rather than described: of
\dsTargets{} targets across \dsFilings{} filings (the \dsLinkedFixPRs{} further targets are
the rows of the per-filing table in \orsec{I} reading ``the fix for the report above''),
\dsMerged{} are merged,
\dsClosed{} are issues each closed when a pull request addressing it merged, not all of those pull requests ours,
\dsOpen{} remain open, \dsEngaged{} drew a comment from somebody other than the
filer, \dsFilerOnly{} carry only the filer's own follow-up, \dsSilent{} nothing
at all, and the
oldest is \dsAgeMax{} days old, elapsed time and not a verdict, since response in
these projects runs to weeks. That tally is the whole disclosure record, and it
carries less than its size suggests: \dsEngaged{} of \dsTargets{} is
a rate at which a report is engaged with and not a rate at which a defect is conceded.
\orsec{I} says which merges repair arithmetic rather than prose, and what a
window of \dsAgeMax{} days can and cannot carry.

Fixes this study did not write have since reached two of the libraries it measures,
and they belong to their author, GitHub user \code{nicoloangileri}. A pull request
moving \code{darts}' conformal models onto the finite-sample order statistic merged
(\#3227) and shipped in release 0.48.0; its \code{statsforecast} counterpart merged
(\#1246) and is not yet in a release; its \code{neuralforecast} counterpart (\#1628)
was closed unmerged, the maintainers moving conformal intervals into a separate
package. Every per-helper row here is a statement about the pinned release, and a
caller on a later one should re-run the conformance suite rather than read the row.

\section{Threats to validity}
\label{sec:limitations}

Each threat names the claim it bounds and its size. \orsec{M} gives each in full, with
the probe that measures it.

\label{sec:construct}%
\textbf{Construct.} Base models are deliberately simple, last-value and linear
forecasters and ridge and logistic regression; a gradient-boosted forecaster beside the
naive one averages \fmDeficitMean{} over its deficit cells (\orsec{E}). The classifier
of \S\ref{sec:suite} has no precision or recall, because the label it would be scored
against is the rank it reads. \textbf{External.} The population is Python packages at
the versions of Table~\ref{tab:census}, and \fsOutside{} conformal-first distributions sit
outside the frame (\orsec{F}). \textbf{Conclusion.} Nesting excludes a sign-symmetric
null, so each forecasting cell's sign-flip test is a check that the cell has enough
discordant units, not a test of an effect: under Holm--Bonferroni \muSurvive{} of
\muFamily{} cells reach $0.05$, and the \muDropped{} that does not is the smallest
difference, \muDroppedDelta{} (\orsec{H}). \orsec{M} gives each threat in full.

\subsection{Researcher bias}
\label{sec:bias}

A sole author wrote the instrument, chose what it would be pointed at, and read its
output \citep{ralph2021}. The \rqOneResisted{} sites that resisted classification by
reading, named in \orsec{M}, were settled by running them through the suite of
\S\ref{sec:suite}, which we wrote. Three things stand in for independence and none of
them is independence: classification by execution against a named candidate set, so a
helper matching nothing is reported as unmatched; an R port of the index arithmetic,
cross-checked over \muCells{} cells; and census anchors re-checked against the
pinned source. Parts of the analysis and gate code were written with a language model's
assistance. The gates compare every printed value with the committed probe output it
came from, which bounds what that code can get wrong in the prose; the R port re-implements
the index arithmetic only, not the rest of the analysis. No maintainer has confirmed a classification. The \cenAbsent{} +
\cenPvalue{} sites whose branch letter is read from source, and the filing rule of
\S\ref{sec:checks}, are the author's judgement alone. \S\ref{sec:conclusions} reports
the one second reading returned so far.

\subsection{Internal validity}

\textbf{Absolute coverage on the raw archive is not attributable to the convention.}
Real series are not exchangeable, so on the archive only the paired difference carries
a claim. The attributable arm of \S\ref{sec:attributable} resamples a series' own
residuals i.i.d.; at the shipped default it delivers \atDefaultGot{} against a
requested \atDefaultNominal{}, inside the bracket the landed index fixes. It does not
reach \code{sktime}'s public path, whose misalignment (\S\ref{sec:align}) breaks
exchangeability whatever the data. \textbf{The index rule misses one library's cells}:
all \robSkCells{} whole-archive \code{sktime} cells fall short of pred.\ $\Delta$ by
more than two clustered standard errors, while every feasible \code{statsforecast} cell
sits within two. \textbf{One shortfall is not the convention at all}: the largest cell
is infeasibility and one further cell is in-sample residual bias, both reported apart.
\textbf{The environment is part of the measurement}: every per-helper row is one
release in one pinned environment, and \code{fortuna}'s clamp above a level of $1$ is
the pinned \code{jax}'s. \textbf{Some choices were made after seeing output}, among
them the coincidence-band grid and the tabular grid's added size, at which every
nonzero single-rail cell of Table~\ref{tab:tabular} sits; \code{PREREGISTRATION.md}
names each.

\section{Discussion and conclusions}
\label{sec:conclusions}

\paragraph{The three questions, and the record.}
RQ1: reading the source does not establish the rule, and here it had to be settled by
running it. \cenPTwo{} sites across \cenLibraries{}
packages, \cenDriven{} of them put in a class by running the site and reading back
its threshold and \cenShared{} by running the \code{numpy} method they share
(\S\ref{sec:census}); \rqOneResisted{} resisted classification by reading and
\rqOneContradicted{} of those \rqOneResisted{} contradicted what their source suggested
(named in \S\ref{sec:bias}). \rqOneResisted{} of
\cenPTwo{} is an existence result, not a rate.
RQ2: the coverage a caller receives does not match the level requested, and the
$(n+1)/n$ correction does not predict the gap. The virtual index does on the
\code{statsforecast} and \code{darts} cells, and on \code{sktime}'s only on M1: on the
whole of M3 all \robSkCells{} \code{sktime} cells fall short of pred.\ $\Delta$ by more
than two clustered standard errors, on M1 \robMOneSkBeyond{} of \robMOneSkCells{} do, and aligning the
residuals does not remove the M3 miss (\S\ref{sec:robustness}). The widest
cell we measure at nominal $0.90$ delivers \sfDefaultNinety{} under a library
default infeasible at that calibration size and \sfMThreeDefaultNinety{} for the same
default on the second archive, while the tabular arm is a null for undercoverage in all
\tabConfigs{} implementations (\S\ref{sec:tabular}). RQ3: below the floor
\confRouteSilent{} of \confPaths{} public call paths hand back a finite number and
say nothing (\S\ref{sec:boundary}), and one documented public keyword whose stated
purpose is to permit an infinite bound skips the guard instead, delivering
\clipDelivered{} against \clipLevel{} at $n = \clipN{}$ on \clipReaches{} regressor
classes (\S\ref{sec:unbounded}). That region is not a corner these projects avoid:
\rfcBelowFloor{} of the \rfcFixtures{} scorable calibration sizes they pin in their
own test suites sit below the floor at $0.90$, so the sizes their tests exercise are
mostly sizes at which no rank carries $0.90$ (\S\ref{sec:checks}).

\paragraph{For a practitioner, two questions replace the one usually asked.}
Not which level was requested, but which rank it resolved to; and whether any rank
could have carried it at the calibration size in hand. The second is the floor
$n \geq 1/\alpha - 1$, and in a multi-step helper it multiplies by the horizon: a
weekly horizon of a year, $H=52$, needs \hzFloorFiftyTwoNinetyFive{} calibration windows at nominal
$0.95$ (\S\ref{sec:horizon}). Neither question is
answerable from the documentation of any package we audit. \S\ref{sec:suite} answers
the first by execution and asks, in a reporting checklist, that a library state the
level it resolved rather than the level it was handed.

\paragraph{What this audit leaves open.} A prevalence rate for the defect across the
ecosystem, which \orsec{F} says why we do not offer; maintainer response at a window
longer than \dsAgeMax{} days; and any library outside Python. How often a reading of a
site agrees with running it is measured by a reading arm that has so far returned one
sheet, from a reader outside the project, a data engineer the author asked personally, who
worked unpaid and did not know the taxonomy: \scCorrect{} of \scScoreable{} scoreable items
labelled as execution labelled them, Cohen's kappa \scKappa{}. That is one coder, not a
reliability coefficient (\orsec{M}).

\subsection{What follows, for whom}
\label{sec:implications}

The threats above bound what these numbers cannot carry; Table~\ref{tab:implications}
gives what they can. Every row rests on a cell of this paper, not an extrapolation from
one; the one figure measured on i.i.d.\ draws rather than an archive, the width price
in the first row, says so, and the last column names where to check it.

\begin{table}[p]
\centering
\caption{What a caller can do about each finding. The action is bounded by the
evidence beside it: nothing here generalises past the pinned versions of
Table~\ref{tab:census}, and the arithmetic in the first two rows is the only part
that holds independently of any release}
\label{tab:implications}
\small
\begin{tabular}{@{}p{0.30\textwidth}p{0.38\textwidth}l@{}}
\toprule
What was measured & What to do about it & Where \\
\midrule
An uncorrected level through linear interpolation delivers at no $n \le
2000$ & Ask for a rank. \code{sort(scores)[k-1]} at $k = \reqrank$ was exact in
every executed cell; \code{inverted\_cdf} at the corrected level is exact in exact
arithmetic and one rank long in \rcExecMiss{} of \rcExecCells{} cells, each where
$(1-\alpha)(n+1)$ is an integer and that level is not representable as a double. On i.i.d.\ draws the switch costs \wdCorrMean{}
times the shipped default's interval width on average (Monte Carlo s.e.\ \wdCorrMeanSe{}) over \wdCells{} i.i.d.\ cells
(\wdOffCorrMean{} over \wdOffCells{} more at sizes not divisible by ten)
(\orsec{H} prices the same switch on the archive, where a coincidence cell
costs nothing and a deficit cell costs \daWidthRatioMean{} times arm A's width per series on average), and
\wdCorrWorst{} at its worst, where the required rank is the sample maximum &
\S\ref{sec:background}, \orsec{C} \\
\addlinespace
No rank carries the level below $n = 1/\alpha - 1$, and a Bonferroni split over a
horizon $H$ raises that floor to $H/\alpha - 1$ & Check $n$ against the floor before
calibrating: \rmTheoreticalNinety{} at $0.90$, \rmTheoreticalNinetyFive{} at
$0.95$, and \hzFloorTwelveNinetyFive{} windows at $H{=}12$ and $0.95$ &
\S\ref{sec:horizon} \\
\addlinespace
\code{sktime}'s \code{conformal\_bonferroni} sits below the horizon-scaled floor in
\hzInfeasCells{} of the \hzCells{} cells measured and returns a finite band in every
one & Check $n \ge H/\alpha - 1$ before requesting a simultaneous band, or request
per-step bands and say so & \S\ref{sec:horizon} \\
\addlinespace
\code{statsforecast}'s \code{ConformalIntervals} defaults to two calibration
windows, below the floor at both levels this paper runs, $0.90$ and $0.95$ & Set \code{n\_windows} above the
floor. The default delivers \sfDefaultNinety{} against $0.90$ on one archive and
\sfMThreeDefaultNinety{} on the other & \S\ref{sec:realdata} \\
\addlinespace
\code{mapie}'s \code{allow\_infinite\_bounds} skips the guard it documents, on
\clipReaches{} public regressor classes & Do not rely on it below the floor; the
flagged path returns \clipDelivered{} against \clipLevel{} at $n = \clipN{}$ where
the default path raises & \S\ref{sec:unbounded} \\
\addlinespace
\code{sktime}'s \code{empirical} spans a median \skEmpNinetyTwentyRank{} gaps at a
median $n = \skEmpNinetyTwentyN{}$ where a median \skEmpNinetyTwentyReq{} are required, and
calibrates a step-$\ell$ forecast on $(\ell{+}1)$-step residuals & Read the rank the call
resolved rather than the level requested. Both methods measured here,
\code{empirical} and \code{conformal}, omit the $(n{+}1)$ correction, so neither of
those removes the deficit & \S\ref{sec:realdata},
\S\ref{sec:horizon} \\
\addlinespace
\code{darts} reaches the required rank on a residue class in $n$ and misses it by
one rank off that class & Pick \code{cal\_length} on the delivering class for the
level in use, or index the rank directly. Collecting more data is a coin flip whose
bias is $2\alpha$, the density item~\ref{itm:residue} of \S\ref{sec:imported} gives
for \code{higher} & \S\ref{sec:severity} \\
\addlinespace
\confRouteSilent{} of \confPaths{} public call paths return a finite number below
the floor and say nothing & Assert feasibility at the call site. Switching library
at one $n$ and level can return an exception, an infinity or a finite number &
\S\ref{sec:boundary} \\
\bottomrule
\end{tabular}
\end{table}

\paragraph{For a maintainer, four of these are cheap.}
Choose test fixtures from the sizes certified for the level under test (which pins the
helper among the candidate rules and no further), not from round numbers: a fixture
pinned to a coincidence cell passes forever
while the helper is wrong at the next $n$ (\S\ref{sec:certificate}), coincidental
correctness \citep{masri2014} built into the fixture. Adopt a
contract for the floor and document it: raising and returning $\pm\infty$ are both
defensible and the ecosystem currently does both, plus a third thing that is
neither. Run the conformance suite against the public entry point rather than the
internal helper, because \cenPThree{} entry points is the surface a caller has. And
where a public keyword promises an infinite bound, test the promise at a size below
the floor, which is the only regime in which it means anything.
\section*{Reproducibility statement}
\label{sec:repro}

\deanon{All probes, committed probe outputs and pinned environments are in the archived
artifact the data availability statement names.}{The reproduction artifact's archival
DOI is withheld for review and inlined at camera-ready.}
Every measured quantity in this paper, in prose, tables and figures, is read from a
committed probe output rather than typed. A ledger, sent with the journal submission and
available from the author, lists every number the paper prints and, where exactly one output carries it, that file and
the first line in it that does; a row locates a value and does not by itself establish
that the line is the computation the value came from. Three rows to check first:
\sfMThreeDefaultNinety{} (a shipped default on the second forecasting archive),
\atDefaultGot{} (the same default on the attributable arm), and $\tabWorstDelta{}$ (the
worst paired difference in the tabular arm).
\deanon{%
The artifact also carries \code{PREREGISTRATION.md}, a written separation of the
decisions fixed before any measurement from those taken after seeing output, with the
counterfactual for each of the second kind. It is not a pre-registration and says so:
only the commit dates settle when each probe appeared. It names the occasions on which
a late choice moved a printed number. The artifact is archived at
\code{10.5281/zenodo.21799203}, the all-versions DOI, which resolves to whatever is
newest; \code{10.5281/zenodo.24287593} is \code{v1.5.15}, the release carrying the
outputs this paper reads.
}{}

\section*{Data availability}

All probes, committed probe outputs and pinned environments are archived at
\code{10.5281/zenodo.21799203}, the all-versions DOI. The release carrying these outputs
is \code{v1.5.15}, at \code{10.5281/zenodo.24287593}. Apart from one volunteer's
classification sheet (\S\ref{sec:conclusions}), archived unedited in the artifact, no data
were collected from people. The forecasting series are the M1, M3 and M4 collections of the Monash
Time Series Forecasting Archive \citep{godahewa2021}, each distributed on Zenodo under
version four of the Creative Commons Attribution licence, as the licence field of every
record states (\code{10.5281/zenodo.4656159}, \code{4656154}, \code{4656298},
\code{4656262} and \code{4656548}). The tabular datasets are the members of OpenML-CC18 and OpenML-CTR23
that the selection rule of \S\ref{sec:subjects} kept, fetched through OpenML under the
licence each dataset's own record states: OpenML's \code{Public} label or the same Creative
Commons Attribution licence for every dataset but one, and a non-commercial scholarly and
teaching licence for \code{socmob}. Nothing from either source is redistributed, and a
probe in the artifact reads every record on a stated date and keeps what it found. The
audited packages are
third-party open-source releases, read at the pinned versions named in the manifest and
redistributed nowhere.

A reviewer who wants to check the paper rather than rebuild it needs one environment and
one command: \code{make verify} re-derives, in under ten minutes, the feasibility floor,
the ranks \code{inverted\_cdf} and \code{higher} reach at the corrected level,
\code{numpy}'s default coverage at one size, and the arithmetic of two below-floor cells,
\S\ref{sec:realdata}'s largest and \S\ref{sec:unbounded}'s, and prints each beside the
committed value it was read from. The full
reproduction needs three pinned environments for the census and coverage arms and one
per out-of-census package, because the audited libraries do not agree on
\code{numpy} and \code{pandas} versions; the manifest names which harness needs which,
and the deposit records those three as freezes, two from the interpreters that produced
the outputs and one reconstructed and validated by re-running its probes; the
out-of-census environments are not recorded.

\deanon{%
\section*{Acknowledgements}
The author thanks Devarsh Patel, who labelled the blind classification sheet
(\S\ref{sec:conclusions}) without knowing the taxonomy it tests.
}{}

\section*{Declarations}

\textbf{Funding.} No funds, grants, or other support were received.

\textbf{Competing interests.} The author has no financial interests to declare. Two
non-financial matters are disclosed in full. First, and directly relevant to the method:
during this work the author made \dsFilings{} filings against the packages audited
here (defect reports, pull requests, and one review of another contributor's pull
request), listed in
\orsec{I}'s disclosure table, one row per target in
the per-filing table of \orsec{I}, which also carries tracker state and age at \dsAsOf{}.
Of the \dsTargets{} targets that table tracks, \dsMerged{} have been merged, one of them
a third party's pull request the author reviewed and did not file. The author is not a
maintainer of, and has no affiliation with, any audited package, and no maintainer of an
audited package is an author of this paper. Second, the author is employed as a data analyst
at Washington University in St.\ Louis. That employment is unrelated to this work, did not
fund or sponsor it, and the university has no interest in its publication; the research was
conducted independently, which is what the affiliation line records.

\textbf{Ethics approval.} Not applicable. This study analyses released software and public
issue trackers. No animals were involved. One volunteer, asked personally by the author and
unpaid, labelled \scDrawn{} code excerpts from released packages (\S\ref{sec:conclusions}); the
task concerned the code, not the person, and no personal data were recorded. No ethics
review was sought.

\textbf{Consent.} The volunteer agreed to complete the sheet, to its archiving in the
artifact, and to being named in the Acknowledgements.

\textbf{Data, materials and code availability.} See the data availability statement above.

\textbf{Author contributions.} Sole author. Shivam Lalakiya conceived the study, designed and
ran all measurements, filed the upstream reports, and wrote the manuscript.

\textbf{Generative AI.} Generative AI (Claude, Anthropic) assisted with drafting and editing
this manuscript, and with parts of the build, gate and analysis code, in every case under the
author's direction and review; the author is accountable for all of it. It did not choose which
packages to audit, what to report, or what to file upstream. One review pass by a language model
pointed to \scAddedAfter{} resolution sites the census had missed; the author confirmed each
against the pinned source before adding it. No measurement in this paper was produced
by a language model: every number is generated from a committed probe output by the build
described in the \emph{Reproducibility statement}, which fails rather than typesetting if a value in the prose
disagrees with the output it came from, or was formed by scaling or dividing two of its values
instead of read from one.
The probes, the archived outputs and that build are in the artifact above, so this statement is
checkable rather than asserted.

\clearpage
\section*{Online Resource 1}
\label{sec:onlineresource}

The parts listed below ship with this submission as Online Resource~1 rather than as
appendices, because they are what a short review may skip, and \S A to \S C carry
the proofs of the results \S\ref{sec:imported} states. Every claim the article
makes is stated in the article, and with the figure it rests on except the
group-conditional amplification, which \S D alone measures; what \S H to \S K carry is the table,
the statement or the narrative behind a claim the article makes in a sentence.

\begin{center}
\begin{tabular}{@{}ll@{}}
\toprule
Online Resource~1 & What it carries \\
\midrule
\S A & The fractional rank, proved here \\
\S B & The tail law, proved here \\
\S C & The delivering set: densities, periodicity and the certificate, proved here \\
\S D & Where the one-rank error stops being $O(1/n)$ \\
\S E & The frequency and the base model \\
\S F & The sampling frame and version history, measured \\
\S G & What the randomised bound delivers, and what it costs \\
\S H & Delivered coverage: replication, attribution and robustness \\
\S I & How the numbers were checked \\
\S J & The census, site by site \\
\S K & Three further results the article uses once each \\
\S L & Terms, in full \\
\S M & Threats to validity, in full \\
\S N & Related work, in full \\
\S O & Every section's claim, and what would have sunk it \\
\bottomrule
\end{tabular}
\end{center}

{\small\setlength{\bibsep}{1pt plus 0.3pt}
\bibliography{refs}}

\end{document}